\documentclass[aps,prd,twocolumn,superscriptaddress,floatfix]{revtex4-1}
\usepackage{axodraw2}
\usepackage{color}
\usepackage{amssymb}
\usepackage[normalem]{ulem}
\usepackage{graphicx,color}
\usepackage{amsmath}
\usepackage{subfigure}
\usepackage[normalem]{ulem}
\usepackage{booktabs}
\usepackage{xcolor}
\usepackage{soul}
\usepackage{epsfig,graphics,color,graphicx,amsmath}
 \usepackage{epstopdf}
  \usepackage{epsfig,graphics,color,graphicx,amsmath}

\def\la{\langle}
\def\ra{\rangle}

\def\be{\begin{equation}}
\def\ee{\end{equation}}
\def\bea{\begin{eqnarray}}
\def\eea{\end{eqnarray}}

\colorlet{darkgreen}{green!50!black}
\colorlet{brightyellow}{yellow!75!red}
\colorlet{orange}{red!50!yellow}
\colorlet{darkblue}{blue!60!black}
\colorlet{darkred}{red!80!black}
\usepackage{array}
\usepackage{array,etoolbox}
\usepackage{fancyhdr}
\usepackage{slashed}

\newcommand{\bno}{\begin{eqnarray*}}
\newcommand{\eno}{\end{eqnarray*}}

\newcommand{\bl}{\begin{large}}
\newcommand{\el}{\end{large}}
\newcommand{\bla}{\begin{Large}}
\newcommand{\ela}{\end{Large}}

\newcommand{\ede}{{\end{document}}}

\def\la{\langle}
\def\ra{\rangle}

\def\be{\begin{equation}}
\def\ee{\end{equation}}
\def\bea{\begin{eqnarray}}
\def\eea{\end{eqnarray}}

\begin{document}

\vspace{-1.5cm}
{\tiny .}
{\normalsize \bf \hspace{10.0ex}  LFTC-24-07/90}  
\vspace{1.005cm}

\title{Off-shell pion properties: electromagnetic form factors 
	and light-front wave functions }

\author{Jurandi Le\~ao}
\affiliation{\em Laborat\'orio de F\'\i sica Te\'orica e Computacional-LFTC, 
Universidade Cruzeiro do Sul / Universidade Cidade de S\~ao Paulo,
015060-000, S\~ao Paulo, SP, Brazil}

\affiliation{Instituto Federal de S\~ao Paulo, 11665-071,  Caraguatatuba, SP, Brazil}

\author{J. P. B. C. de Melo}
\affiliation{\em Laborat\'orio de F\'\i sica Te\'orica e Computacional-LFTC, 
Universidade Cruzeiro do Sul / Universidade Cidade de S\~ao Paulo,
015060-000, S\~ao Paulo, SP, Brazil}

\author{  T. Frederico}
\affiliation{\em Instituto Tecnol\'ogico de
Aeron\'autica, 12.228-900 S\~ao Jos\'e dos Campos, SP,
Brazil}

\author{ Ho-Meoyng Choi}
\affiliation{\em Department of Physics, Teachers College, Kyungpook National University,
     Daegu, Korea 41566}

 \author{Chueng-Ryong Ji}
\affiliation{\em Department of Physics, North Carolina State University,
Raleigh, NC 27695-8202, USA}

\begin{abstract}
The off-shell pion electromagnetic form factors are explored with 
 corresponding off-shell light-front wave functions modeled 
by constituent quark and anti-quark. We apply the Mandelstam approach
for the microscopic computation of the form factors relating the 
model parameters with the pion decay constant and charge radius. Analyzing  
the existing data on the cross-sections for the Sullivan process, 
$^1$H$(e,e',\pi^+)n$ \cite{Blok2008}, we extract 
the off-shell pion form factor using the relation derived from the 
generalized Ward-Takahashi identity for the pion electromagnetic current.
 They are compared with our previous results~\cite{Choi2019}
from exactly solvable manifestly covariant model of a (3+1)-dimensional fermion field theory. 
 We find that the adopted constituent quark model reproduces the extracted off-shell form factor $F_1(Q^2,t)$
	 from the experimental data~\cite{Blok2008} 
within a few percent difference and matches well with our previous theoretical
simulation~\cite{Choi2019} which exhibits a variation of about 10\% for 
the extracted off-shell pion form factor $g(Q^2,t)$. We also identify
the pion valence parton distribution function (PDF) and transverse momentum distribution (TMD)
in terms of the light-front wave function and discuss their off-shell properties.
\end{abstract}
\date{\today}
\maketitle

\section{Introduction}

The pion plays a special role in hadron physics as the Goldstone boson of the dynamical 
chiral symmetry breaking in the light  SU(2) flavor sector within 
Quantum Chromodynamics (QCD) while it is the ground state of the 
pseudoscalar combination of the $q\bar{q}$ system, see e.g.~\cite{Aguilar2019}.
 Because the valence structure of the pion is rather simple as a quark-antiquark two-body system, 
it provides a good testing ground for 
	 QCD inspired models along with dynamical predictions
	  from lattice QCD simulations and other Euclidean approaches~\cite{Roberts:2021nhw}.
 In particular,  the electromagnetic (EM) form factor $F_{\pi}(Q^2)$ 
 of the pion provides information on the pion structure depending on
 the four-momentum squared $q^2(=-Q^2)$ of the virtual photon.
 While the pion form factor for low $Q^2$ region can be measured directly from the elastic scattering of pions off atomic electrons~\cite{Dally1,Dally2,Amen1,Amen2,Baldini2000},
it is hard to extract the form factor for intermediate and high $Q^2$ regions 
directly from the elastic scattering experiment as the pion cannot be an amenable
	 target due to its short lifetime.

One method of indirect experimental extraction on the pion form factor for the intermediate
 and high $Q^2$ regions comes from the exclusive version of the
Sullivan process~\cite{Sull}, namely,
from the 
cross sections of the pion electroproduction reaction
$^1$H$(e,e'\pi^+)n$~\cite{Blok2008,Huber2008,Thorn1,Thorn2,Tadevosyan2007}. In the exclusive Sullivan process~\cite{Blok2008,Huber2008,Thorn1,Thorn2,Tadevosyan2007}, 
a  virtual pion with an  off-shell mass ($p^2=t <0$) from the  proton's cloud
absorbs a virtual photon with space-like four-momentum $q (=p'-p)$
 transferred by the electron, and becomes an on-mass shell pion ($p'^2 = m^2_\pi$).
However, due to the nature of the virtual pion target, the form factor obtained
 from this reaction is the off-shell EM form factor.  

In general, the off-shell EM vertex of the pseudoscalar meson
requires~\cite{Nish,Rudy1994,Weiss94,Choi2019,Choi2020} 
two form factors $F_1$ and $F_2$. 
These form factors depend on the invariant masses associated with the initial and final pion momenta 
in addition to the squared momentum transfer. 
The two form factors are related by the generalized Ward-Takahashi identity (WTI)~\cite{Ward50,Taka57}.

Accordingly, to extract the on-shell ($t=m^2_\pi$) pion EM form factor from
 the off-shell form factor via
$F_{\pi}(Q^2)=F_1(Q^2, t=m^2_\pi)$,
the reaction $^1$H$(e,e'\pi^+)n$ is anticipated  to be dominated by small values of
$t$ in the proton to the neutron transition, 
as the pion pole is in the timelike region
$t=m^2_\pi > 0$ of the $t$-channel ~\cite{Blok2008,Huber2008}.
 Thus, it is crucial to isolate the longitudinal cross section $\sigma_L$
that encapsulates the meson exchange process for a clean access to the pion off-shell longitudinal 
form factor $F_1(Q^2,t)$ by minimizing background contributions~\cite{Blok2008,Huber2008}. 
The separation of differential cross section according to the polarization states of the virtual photon is known as the Rosenbluth separation and it is expressed in terms of the
 longitudinal differential cross section ($d\sigma_{\rm L}/dt$), 
the transverse differential cross section ($d\sigma_{\rm T}/dt$), and two other
 differential cross sections resulting from interference 
($d\sigma_{\rm LT}/dt$ and $d\sigma_{\rm TT}/dt$).

The transverse form factor $F_2(Q^2,t)$ can, in principle,
be obtained from $F_1(Q^2,t)$  according to the WTI for 
the pion off-shell EM current~\cite{Rudy1994}. 
However, the knowledge of $F_1(0,t)$ is required to obtain $F_2(Q^2,t)$
 while it is hard to be obtained experimentally. Thus, one may resort to models
  to obtain this quantity.
In Ref.~\cite{Choi2019}, we have computed $F_1(0,t)$ using a manifestly
covariant quantum field theoretic model of the pion Bethe-Salpeter (BS) 
amplitude with a structureless constant pion-quark vertex. We incorporated constituent quark degrees
 of freedom based on the Mandelstam approach to the microscopic pion EM current. 
In that analysis~\cite{Choi2019}, we also found a new measurable form 
factor in the on-shell limit by defining $g(Q^2,t)=F_2(Q^2,t)/(t- m^2_\pi)$.

 Especially, we showed that $g(Q^2=0, t=m^2_\pi)$ is related with
  the pion charge radius via $g(0, m^2_\pi)=\langle r^2_\pi\rangle/6$.
The extraction of the pion EM form factor from the reaction $^1$H$(e,e'\pi^+)n$
 also needs a model for the pion-nucleon pseudo-scalar form factor, 
namely $G_{\pi NN}(t)$~\cite{Machleidt1987,Choi2019,Choi2020}. 
However, a slight dependence on $G_{\pi NN}(t)$ in extracting the
	 pion EM form factor from $d\sigma_{\rm L}/dt$ can generally be
	  disregarded at $t\lesssim (0.2-0.3)\,$GeV$^2$, where the 
	  cross-section data is typically gathered~\cite{Blok2008}.
It should be mentioned that the possible off-shell effects
also appear in the pion exchange current as discussed in Ref.~\cite{Gross1987}. 
Some recent  works have investigated  the off-shell effects 
on the pion EM form factors within the BS approach~\cite{Quin2018}, 
and the analyses in Refs.~\cite{Perry2019,Perry2020} 
utilized the Vanderhaeghen, Guidal, and Laget Regge model~\cite{Guidal1998}.

In the present work, we go beyond the previous quantum field theoretic constant (CON)
 vertex study ~\cite{Choi2019} to explore the more phenomenological relativistic constituent quark  
 model~\cite{deMelo1999,deMelo2002}  in the light-front (LF) framework for  
 the analysis of the pion off-shell EM form factors, $F_1$ and $F_2$. 
In contrast to the previous manifestly covariant field theoretic model
 computation utilizing the dimensional regularization, 
 the present LF constituent quark model 
uses the Pauli-Villars regularization to identify 
the LF quark model wave functions involved in the half-off shell pion form factors. 
We adopt a symmetric (SYM) ansatz for the pion BS amplitude with a pseudoscalar coupling
 of the constituent quark and antiquark to the pion field and find the corresponding
  on-shell and off-shell LF wave functions for the computation of the half-off shell
   pion form factors. We compute the pion decay constant as well as the half-off shell pion
    form factors to fix both the constituent quark/antiquark mass and the Pauli-Villars regularization mass consistent with experimental data. We compare the results obtained in this work (SYM) with 
those obtained previously (CON),
 denoting them as SYM and CON respectively. We also discuss
  the LF wave function-related quantities such as the pion valence
   parton distribution function (PDF) and transverse momentum distribution (TMD).

This work is organized as follows. 
In Sec.~\ref{sec:basics}, we briefly review the basic formalism. We define the pion off-shell EM 
form factors in Sec.~\ref{sec:off-shell} and provide the cross-section
 formulas for the pion electroproduction
  process $^1{\rm H}(e,e'\pi^+)n$ in Sec.~\ref{subsec:xsection}.
The microscopic pion EM current, based on the Mandelstam approach, is detailed in
Sec.~\ref{sec:miccurrent}. We present our SYM model in Sec.~\ref{subsec:models},
 and the corresponding on-shell and off-shell pion wave functions
  in Sec.~\ref{subsec:comptec}. In Sec.~\ref{results}, we present 
  our results for the off-shell form factors (Sec.~\ref{subsec:offshelff})
   and the cross-sections  (Sec.~\ref{subsec:xsec}).
  We then conclude our work in Sec.~\ref{conclusions}.
In Appendix~\ref{app:kmint}, we demonstrate that the loop integration in the LF energy of the 
Mandelstam formula for the plus component of the pion EM  current is free 
of LF zero modes in the Drell-Yan frame. In Appendix~\ref{app:tables}, we present
the tables of the form factors and cross-sections for further details of our results.

\section{Basics}
\label{sec:basics}

\subsection{Off-shell EM form factors} 
\label{sec:off-shell}

In what follows, we give a brief discussion on the pion off-shell EM vertex following closely Refs.~\cite{Rudy1994,Choi2019,Broniowski:2022iip} to introduce the notation and the main
 quantities calculated in our work. We start with the pion-photon vertex, $\Gamma^\mu$,  
 written in a general manner in terms of  two off-shell form factors
due to the pseudoscalar nature of the pion:
\begin{eqnarray}
\label{eq1}
\Gamma^\mu(p',p)=e [ P^\mu G_1(q^2,p^2,p^{\prime 2}) + 
q^\mu G_ 2(q^2,p^2,p^{\prime 2})] , \,\,\,\,
\end{eqnarray} 
where 
 $e$ is the electric charge, $p^{(\prime)\mu}$ is the pion initial
  (final) momentum, $P^\mu=(p'+p)^\mu$, and $q^\mu = (p' - p)^\mu$ 
is the momentum transfer. The pion charge normalization is  $G_1( 0,m^2_\pi,m^2_\pi) = 1$
 for the fully on-shell form factor.

The WTI satisfied by the pion-photon vertex~\cite{Rudy1994,Zuber1980} 
provides the following relation for the fully off-shell vertex: 
\begin{equation}
\label{eq2}
q_\mu \Gamma^\mu (p',p) = \Delta^{-1}(p') - \Delta^{-1}(p)~,
\end{equation}
where $\Delta(p)$ is the full renormalized  pion propagator, 
\begin{equation}
\label{eq3}
\Delta(p)~=~\frac{1}{p^2 - m^2 -\Pi(p^2) + \imath \epsilon}\, ,
\end{equation}
and $\Pi(p^2)$ is the renormalized pion self-energy  constrained by
 the on-mass shell condition $\Pi(m^2_\pi)=0$.

From the WTI given by Eq.~\eqref{eq2}, we find that
\bea\label{eq:WTI}
&&(p'^2 - p^2)G_1(q^2,p^2,p'^2) + q^2G_2(q^2,p^2,p'^2) \nonumber\\
&&=\Delta^{-1}(p') - \Delta^{-1}(p)\,. 
\eea
We note that 
the final state pion is on-mass-shell, $p^{\prime 2}=m^2_\pi$, with $\Delta^{-1}(p')=0$ for
the reaction $^1{\rm H}(e,e'\pi^+)n$. 
In addition, for  the case of a real photon $q^2=0$, one finds  from Eq.~(\ref{eq:WTI}) that
\begin{equation}\label{eq:WTI2}
(p^2-m^2_\pi )\,G_1(0,p^2,m^2_\pi)   = \Delta^{-1}(p)\, , 
\end{equation}
and from  Eqs.~(\ref{eq:WTI}) and~(\ref{eq:WTI2}), one has that
\bea
\label{eq:WTI3}
& & G_2(q^2,p^2,m_\pi^2) \nonumber\\
& & = \frac{ (m^2_\pi - p^2)}{q^2} 
\left[G_1(0,p^2,m_\pi^2) - G_1(q^2,p^2,m_\pi^2)\right].
\eea

In this case of the pion initial state being off-mass shell ($p^2 =t$)
 but the final state being on-mass shell ($p'^2=m^2_\pi$), 
we rewrite Eq.~(\ref{eq:WTI3}) as 
\begin{equation}
F_2(Q^2,t) = \dfrac{t - m_\pi^2}{Q^2}
\left[F_1(0,t) - F_1(Q^2,t)\right]\,,
\label{f2ffactor}
\end{equation}
where $F_i(Q^2,t) \equiv G_i(q^2,t,m_\pi^2)$ $(i = 1,2)$ and $Q^2 (= -q^2)$ is the square of the four-momentum
transfer in the spacelike region.  
It can be seen from Eqs.~\eqref{eq:WTI3} and~\eqref{f2ffactor}
that $G_2(q^ 2=-Q^2,m^2_\pi,m^2_\pi)=F_2 (Q^2, m^2_\pi) = 0$ when both the initial and final mesons are on-shell. 
This indicates the conservation of the pion electromagnetic current.

From Eqs.~\eqref{eq1} and~\eqref{f2ffactor}, 
the half on-shell ($p'^2=m^2_\pi$) and  half off-shell ($p^2=t$) pion-photon vertex
is given in terms of $F_1(Q^2,t)$ as follows
\begin{multline}
\label{vertex2}
\Gamma_\mu(p',p)|_{p'^2=m^2_\pi,p^2=t} = \,e  \biggl[
(p'+p)^\mu\,F_1 (Q^2,t)     \\
 +  (p'-p)^\mu \frac{(t-m^2_\pi)}{Q^2}\,(F_1(0,t)-F_1(Q^2,t))
\biggr]\, .
\end{multline} 
The pion charge normalization is expressed by $F_1(Q^2=0,m_\pi^2)=G_1(0,m^2_\pi,m^ 2_\pi)=1$. 
In the electroproduction process, directly measuring the 
form factor $F_2(Q^2, t)$ is impractical due to the transversality 
of the electron current. 
Moreover, $F_2(Q^2, t)$ tends to zero as $t\to m^2_\pi$. 
However, despite this limitation, the ratio of $F_2(Q^2, t)$ to $t - m^2_\pi$ remains nonzero 
when $t$ approaches $m^2_\pi$.

In Ref.~\cite{Choi2019},  the new form factor $g(Q^2,t)$ was defined as
\begin{equation} 
g(Q^2,t) \equiv \dfrac{F_2(Q^2,t)}{t - m_\pi^2} = 
\dfrac{1}{Q^2}[F_1(0,t) - F_1(Q^2,t)].
\label{g2ffactor}
\end{equation}
In addition, it was shown that $g(Q^2=0,m^2_\pi)$,
  i.e. in the on-mass shell limit $t=m^2_\pi$ and at $Q^2=0$,
is related with the charge radius of the on-shell pion EM form factor 
via the relation
\begin{equation}
\label{eq10-cptj}
g(Q^2=0,m^2_\pi) = -\frac{\partial}{\partial Q^2}F_1(Q^2=0,m^2_\pi) =  \frac16 \langle r^2_\pi\rangle,
\end{equation}
and the on-mass shell solution for $g(Q^2,m^2_\pi)$ is given by~\cite{Choi2019}
\bea
g(Q^2,m_\pi^2) = 
\frac16 \langle r^2_\pi\rangle\, +\, 
\alpha Q^2 +\, \cdots \, ,
\label{eq:gqmpi}
\eea
where 
\be
2\alpha= -
\left[\frac{\partial}{\partial Q^2}\right]^2F_1(Q^2,m^2_\pi)\Big|_{Q^2=0}\, .
\ee

We remind that in the elastic electron scattering, the cross section comes 
from the contraction of the pion EM tensor with the leptonic one. 
The second term in Eq.~\eqref{vertex2} carrying  $q^\mu$ when  contracted with
the leptonic tensor vanishes, as the electron current is conserved.
 
In  the exclusive version of the Sullivan process where  an off-shell pion 
is taken from the proton cloud  and turns into the on-shell pion, 
only $F_1(Q^2,t)$ is necessary to compute the cross-section for $t$ close
 to the pion mass pole. 
Our focus will be on computing this form factor and using
 Eq.~\eqref{g2ffactor} to obtain $g(Q^2,t)$, with different models.

\subsection{Cross-section formulas for $^1{\rm H}(e,e'\pi^+)n$}
\label{subsec:xsection}

The cross-section  for the exclusive reaction $^1{\rm H}(e,e'\pi^+)n$
 is written in terms of the conventional  longitudinal (L), transverse (T), and  interference(LT and TT) terms 
 as~\cite{Huber2008,Blok2008}
\begin{equation}
\begin{aligned}
(2 \pi) &\frac{d^{2} \sigma}{d t d \phi} =\frac{d \sigma_{\rm T}}{d t}+\epsilon
\frac{d \sigma_{\rm L}}{d t} \\
&+\sqrt{2 \epsilon(\epsilon+1)} \frac{d \sigma_{\rm L T}}{d t} \cos \phi+
\epsilon \frac{d \sigma_{\rm T T}}{d t} \cos 2 \phi\, ,
\end{aligned}
\end{equation}
where 
\be
\epsilon=\left(1+\frac{2|\mathbf{q}|^{2}}{Q^{2}} 
\tan ^{2} \frac{\theta_{e}}{2}\right)^{-1}
\ee
 is the polarization of the virtual photon~\cite{Blok2008}, $\mathbf{q}$ is its three-momentum, 
and $\theta_{e}$ is the angle between initial and final electron momenta. The Rosenbluth separation allows 
 to separate out the longitudinal cross section $\sigma_{\rm L}$,
  which is chosen to minimize the background contributions to the cross-section.
    
\begin{figure}[b]
\begin{center}
\includegraphics[height=3.8005cm,width=4.8005cm]{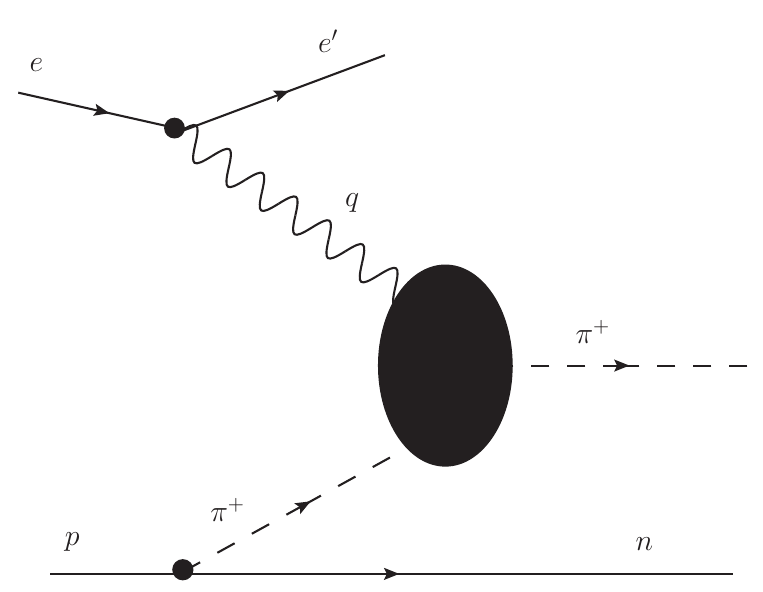}
\caption{
Diagrammatic representation of the pion 
	  pole contribution to $p(e,e')\pi^+ n$ process:
	 The black blob represents the half-on-mass shell photo absorption
	  amplitude. }
 \label{fig:peep}
\end{center}
\end{figure}
The pion photo-production cross-section is calculated with the Chew-Low approach
 for the Sullivan process, that for small values of $t$ is the Born-term  formula dominated by the 
pion-pole contribution illustrated by the Feynman diagram in Fig.~\ref{fig:peep}.
 In this approach,  the pion-pole contribution to $\sigma_{\rm L}$ is given by
\begin{equation}
	N\dfrac{d\sigma_{\rm L}}{dt} = 
	4 \hbar c \left(e G_{\pi NN}\right)^2
	\dfrac{-t Q^2}{(t - m_\pi^2)^2} F_\pi^2(Q^2),
	\label{eq:chew}
\end{equation}
where $e^2/(4\pi \hbar c) = 1/137$ is the fine structure constant and $N$ is a flux
factor~\cite{Blok2008,Choi2019}.   For the form factor $G_{\pi NN}(t)$, we use the typical monopole type of parametrization~\cite{Blok2008,Choi2019} 
\begin{equation}
G_{\pi NN}(t) = G_{\pi NN}(m_\pi^2)
\left(\dfrac{\Lambda_\pi^2 - m_\pi^2}{\Lambda_\pi^2 - t}\right), 
\end{equation}
where  $G_{\pi NN}(m_\pi^2)$ = 13.4 and $\Lambda_\pi = 0.80$\;GeV
 have been taken in the extraction of $F_\pi$ from the Jefferson Lab experiment~\cite{Huber2008}. This
value of $\Lambda_\pi$ is in agreement with the nucleon-nucleon scattering data and also with the deuteron properties~\cite{Machleidt1987,Ericson2002}.

\section{Microscopic Pion EM current}
\label{sec:miccurrent}

\subsection{Model description}
\label{subsec:models}

\begin{figure}[t]
\begin{center}
 \includegraphics[height=20.45cm,width=20.45cm]{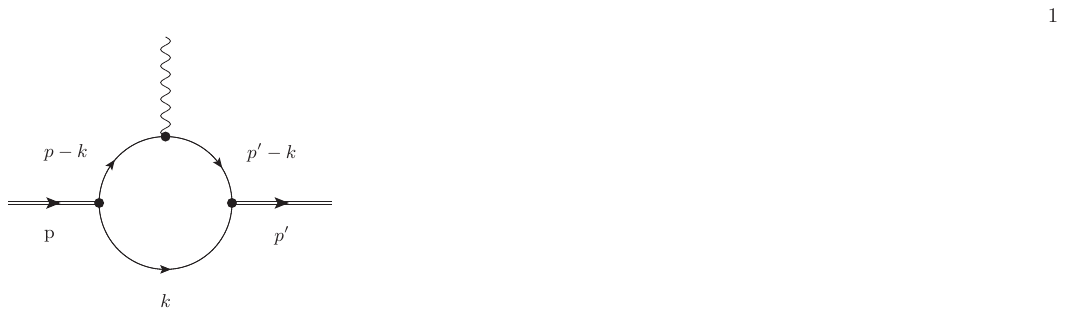}
\vspace{-16.5cm}
\caption{Feynman triangular diagram 
	representing the matrix element of the pion EM current  
	within the Mandelstam framework.}
\label{fig:feydiag}
\end{center}
\end{figure}

 Our model for
the pion microscopic electromagnetic current  is based on
 the Mandelstam amplitude of the photo-absorption
as depicted in
Fig.~\ref{fig:feydiag}, and fulfills the two physical constraints:
 (i) covariance and (ii) current conservation for  the on-mass shell
  initial and final pions. 
Those minimal requirements are sufficient to allow the decomposition 
of the pion-photon vertex  given by Eq.~\eqref{eq1}. 
Furthermore, we adopt the structureless constituent quark picture such 
that the free quark-photon vertex satisfies the WTI, 
 which is essential for the current conservation of 
the Mandelstam representation  for the matrix element of the 
 microscopic pion current operator.

 As in previous applications~\cite{Frederico1992}, we use an 
effective Lagrangian approach with pion and quark degrees of freedom, 
where the coupling of the constituent quark to the pseudoscalar isovector 
pion field within the SU(2) flavor symmetry is given by
\begin{equation}
\mathcal{L} = -i\, \frac{m}{f_\pi}\,\vec{\pi}\cdot 
\bar{q}\,\gamma_5 \,\vec{\tau}\,q\,,
\label{eq:Leff}
\end{equation}
 where $f_\pi$ is the pion decay  constant 
and $m$ is the constituent quark mass. 
 This effective Lagrangian is associated  with a 
point-like  pion-quark vertex, which was indeed adopted in the 
previous study of the pion off-shell EM 
form factors~\cite{Choi2019}.

In the Mandelstam framework of the pion-photon absorption amplitude with a free 
constituent quark-photon vertex, the matrix element  of Eq.~\eqref{eq1} in
the present model is written as
\begin{multline}
\Gamma^\mu(p',p)=
 -2i \, e\dfrac{m^2}{f_\pi^2} N_c 
\int 
\dfrac{d^4 k}{(2\pi)^4}\text{Tr}\Big[S(k)\gamma^5 
S(k - p') \\ 
\times \gamma^\mu S(k - p) \gamma^5 \Big]\Gamma_\pi(k,p')\Gamma_\pi(k,p)\,,
\label{eq:Mandpioncurr}
\end{multline}
where $S(p) = i/(\slashed{p} - m + i\epsilon)$ is the constituent quark propagator,
$N_c=3$ is the number of colors, $q=p'-p$ is the momentum transfer, 
and $k$  is the spectator quark momentum. 
 For the analysis of half off-shell pion form factors from Eq.~\eqref{eq:Mandpioncurr}, we set
the pion in the initial state being off-mass shell
 ($p^2 =t<0$) and the pion in the final state being on-mass shell ($p'^2 = m^2_\pi)$.
Eq.~\eqref{eq:Mandpioncurr} also satisfies the current conservation,
 $q_\mu \Gamma^\mu(p',p)|_{p^2=p^{\prime2}=m^2_\pi}=0$, when both pions are on-mass-shell.

In the previous simple 
covariant model analysis~\cite{Choi2019}, the pion-quark ($\pi q{\bar q}$)
 vertex $\Gamma_\pi(k,p)$ was chosen
to be structureless point-like constant (CON), i.e, 
\begin{equation}
\label{eq:cov}
    \Gamma_\pi^{\rm CON}(k,p)=g_{\pi q\bar q} \, ,
\end{equation}
where $g_{\pi q\bar q}$ is the pion-quark coupling constant. In this model for the point-like vertices,
the fermion-loop was regulated by dimensional regularization 
and the UV divergence
was eliminated by  redefining the renormalized form factor 
as $F^\text{ren}_1(Q^2,t)=1+(F_1(Q^2,t)-F_1(0,m^2_\pi))$.

For the present LF constituent model analysis, 
we regulate the fermion loop using the Pauli-Villars regularization 
methodology to ensure the covariance and the gauge invariance of the model computation.
In particular, we use the symmetric (SYM) vertex to 
smear the $q{\bar q}$ bound-state vertex for the equal mass 
constituents of the pion in a covariant manner~\cite{deMelo2002}:
\begin{small}
\begin{equation} 
\label{eq:sym}
\Gamma_\pi^{(\rm SYM)}(k,p) = N\left(\dfrac{1}{k^2 - m_R^2 + i\epsilon} + 
\dfrac{1}{(p - k)^2 - m_R^2 + i\epsilon}\right)~.
\end{equation}
\end{small}
The Pauli-Villars regularization mass $m_R$ plays the role of
 a momentum cutoff to be fixed by fitting the pion decay constant.


\subsection{Pion wave functions on the light-front}
\label{subsec:comptec}

 The evaluation of the elastic photo-absorption transition amplitude, 
Eq.~\eqref{eq:Mandpioncurr}, is performed in the Drell-Yan 
frame~\cite{Brodsky:1997de} where $q^+=q^0+q^3=0$,
with the choice of the LF energy $(p^-=p^0 - p^3)$ and momentum $(p^+, p_\perp)$ satisfying
\be
\label{eq:22}
p^- =\frac{p^2_\perp+t}{p^+}\, ,
\ee
with $\vec p_\perp=-\vec q_\perp/2$  for the initial state off-mass shell pion, and 
\be
p^{\prime-}=\frac{p^{\prime 2}_\perp+m^2_\pi}{p^+}\,, 
\ee 
with $\vec p^{\,\prime}_\perp=\vec q_\perp/2$ for the final state on-mass shell pion.
The final state on-shell pion  momentum component is chosen as  
$p^{\prime +}=p^{\prime -}$, which defines the kinematics, both
for $t<0$ and $t=m^2_\pi$ in the initial state pion and the final
 state pion, respectively. By satisfying the LF energy-momentum dispersion
  relation described in Eq.~\eqref{eq:22}, 
  we can derive the off-mass shell pion LF wave function.

The model for the BS amplitude utilized in the present work is written as
\begin{equation}
\Psi_i(k,p)=\frac{m}{f_\pi} S(k) \gamma^5 \Gamma(k,p)  \tau_i 
S(k-p).
\end{equation}
The pion wave functions for the 
SYM vertex model is derived after the $k^-$ integration in the 
Feynman amplitude for the photo-absorption process, as detailed in~\cite{deMelo1999,deMelo2002}. 

The spin configuration in the SYM vertex model, i.e., trace term in Eq.~\eqref{eq:Mandpioncurr},
is closely related with the Melosh transformation~\cite{choi24} in the LF constituent quark model, which relates the instant Dirac spinor to the LF helicity spinor. 
According to the Melosh transformation, the pion wave function contains $\uparrow\uparrow$ and $\downarrow\downarrow$ LF helicity components, 
in addition to the singlet $(\uparrow\downarrow -\downarrow\uparrow)$ LF helicity component. 
Typically, $\uparrow\uparrow$ and $\downarrow\downarrow$ LF helicity components are associated with the transverse momentum dependent coefficients related to the orbital angular momentum,
 while the singlet $(\uparrow\downarrow -\downarrow\uparrow)$ LF helicity component carries a constant coefficient associated with the constituent quark mass. 
 The contribution of the $\uparrow\uparrow$ and $\downarrow\downarrow$ LF helicity components to the form factor 
 becomes more significant at the higher $Q^2$ ranges due to the transverse momentum-dependent coefficients associated with these triplet components, beyond the singlet component.

 The wave function corresponding to the LF projection of the SYM vertex function given by Eq.~\eqref{eq:sym},
applicable to both on-mass shell and off-mass shell pions, is given by
 \begin{multline}
\Psi(x,k_\perp,t= p^+p^- -p_\perp^2) = 
\frac{{\cal N}}{t -M^2_0(m^2,m^2)}\\
\Bigg[\frac{1}{x\left(t - M^2_0(m^2_R,m^2) \right)} \\
+
\frac{1}{(1-x)\left(t -M^2_0(m^2,m^2_R )\right)} 
 \Bigg]\, , \label{eq:symwf}
\end{multline}
where 
\begin{eqnarray}
	& & M^2_{0} (m_a^2, m_b^2)= 
	\frac{k^2_\perp+m_a^2}{x}+ \frac{(p-k)^2_\perp+m_b^2}{1-x},
\end{eqnarray}
with $m_a$ and/or $m_b$ taken accordingly as the constituent quark/antiquark 
mass $m$ or the Pauli-Villars regularization mass $m_R$.  
Here, the  normalization constant
${\cal N}=N \sqrt{N_c}  \frac{m}{f_\pi}$ 
with the number of colors $N_c=3$, $x=k^+/p^+$
and $f_\pi$ is the weak pion decay constant.
We should note that the wave functions given by Eq.~\eqref{eq:symwf} is on-shell
 when $t =p^+p^--p_\perp^2 =m^2_\pi$, but off-shell otherwise.

The LF wave function satisfies the normalization at $t=p^2=m^2_\pi$:     
\begin{equation}\label{wfnorm}
	\int^1_0 dx  \int \frac{d^2k_\perp}{16  \pi^3}  |\Psi(x,k_\perp)|^2 \ =  \ 1 ~.
\end{equation}
We also note that the twist-2 TMD, $f(x, k_\perp)$, and the PDF, $f(x)$, 
of the on-mass shell pion
can be identified as~\cite{Lorce,LPS16,choi24}
\begin{equation}\label{TMD}
	f(x, k_\perp)=\frac{1}{16  \pi^3}  |\Psi(x,k_\perp)|^2 ,
\end{equation}
and 
\begin{equation}\label{PDF}
	f(x)=\int d^2 k_\perp  f (x, k_\perp),
\end{equation}
respectively. They satisfy the sum rule given by~\cite{Lorce,LPS16,choi24}
\be\label{eq:f1qTMDnorm}
\int dx\int d^2  k_\perp f (x, k_\perp) = \int dx f (x) =1.
\ee
Using the same normalization constant obtained from the on-mass shell pion wave function, we can also 
compute the TMD and PDF of the off-mass shell pion, as we shall show in Sec.~\ref{results}.


 
Now, the off-mass shell EM form factor $F_1(Q^2,t)$, with the final state on-shell pion, can be rather straightforwardly obtained from the plus component 
of the EM current $\Gamma^+(p',p)$ [see Eq.~\eqref{eq1} for $q^+=0$].
This derivation implies using the free constituent quark
current $\gamma^+=\gamma^0+\gamma^3$ in Eq.~\eqref{eq:Mandpioncurr}, 
which gives (see Appendix~\ref{app:kmint} for details):	
\begin{equation}
F_1(Q^2,t)  = \frac{\Gamma^+(p',p)}{2\,e\,p^+}\,.
 \label{ffactor2}
\end{equation}
Once $F_1(Q^2,t)$ is obtained, then Eq.~\eqref{f2ffactor} provides $F_2(Q^2,t)$.
Notably, the use of $\gamma^+$ eliminates the 
instantaneous terms of the LF fermion propagators attached to the 
quark EM current.  
We should note that the form factor includes all spin helicity contributions, 
which are accounted for in the trace term of Eq.~\eqref{eq:Mandpioncurr}, 
with the explicit LF expression provided in Eq. (A3).

In Appendix~\ref{app:kmint}, we describe how the loop 
	integration over the LF energy, $k^-$, in Eq.~\eqref{eq:Mandpioncurr} is analytically performed
for the SYM vertex function in the valence region ($0<k^+<p^+$) as well as
its asymmetric extension applicable to the unequal constituent mass
 system, such as the kaon. Subsequently, we show that the EM form factors obtained in 
the $q^+=0$ frame using the plus component of the current are immune to the
 LF zero-mode contributions, whether the vertex model is 
 symmetric or asymmetric. 
  It should be noted that for the $q^+ \neq 0$ frame,  preserving the
 full covariance of the model requires accounting for the non-valence
  contributions to the form factor,
   as discussed in Refs.~\cite{Bakker:2000pk,deMelo2002}. 
   
    It is important to observe that our computation of the electromagnetic is fully covariant and contains contributions from all the spin components of the pion valence wave function and in addition other terms which coop with the covariant analytic structure of the SYM BS amplitude.

\begin{center}
\begin{table}[t]
	\centering
 \caption{Results 
  for ${\sqrt{\la r^2_{\pi}\ra}}$, $f_\pi$, $g(0,m^2_\pi)$ obtained 
  from three different $\pi q{\bar q}$ vertices, i..e
$\Gamma_\pi (k, p) = (\Gamma^{\rm CON}_\pi, \Gamma^{\rm SYM}_\pi$). 
 The experimental data  are taken from the PDG~\cite{pdg2020}.} 
 \item[]
\begin{tabular}{l|l|l|l}
\hline\hline
 Model   & ${\sqrt{\la r^2_{\pi}\ra}}$ [fm] & $f_{\pi}$ [MeV] &  $g(0,m^2_\pi)$ [GeV$^2$] \\
\hline
${\Gamma_\pi^{(\rm SYM)}}$       &~0.736           &~92.40            &  2.32           \\
${\Gamma_\pi^{(\rm CON)}}$          &~0.713$\pm$0.013  &~--    & 2.18 $\pm$0.08     \\ 
Exp.~\cite{pdg2020}     &~0.672(8)     &~92.28(7) & 1.93(5)                   \\  
\hline\hline
\end{tabular}
 \label{table0}
\end{table}
 \end{center}

\section{Results}
\label{results}

For the previous study of the manifestly covariant quantum field theoretic constant (CON) vertex model, we adopted $m=0.14\,$GeV and $g_{\pi q\bar q}=1.20$, 
which correspond to the median value of the parameter 
interval obtained in~\cite{Choi2019}.
In this work, the model parameters for the SYM vertex model are the mass of 
the constituent quark,  $m=m_u=m_d=0.22\,$GeV, and a regulator mass 
$m_R=0.6\,$GeV~\cite{deMelo2002} for a consistent description 
of the pion decay constant $f_\pi \approx 92$MeV. 
In Table~\ref{table0}, we present the values of the pion charge radius
 $r_\pi$ and $g(0,m^2_\pi)$ as obtained from Eq.~\eqref{eq10-cptj} in addition 
 to the decay constant $f_\pi$.

 \begin{figure}[t]
 	\begin{center}
\epsfig{figure=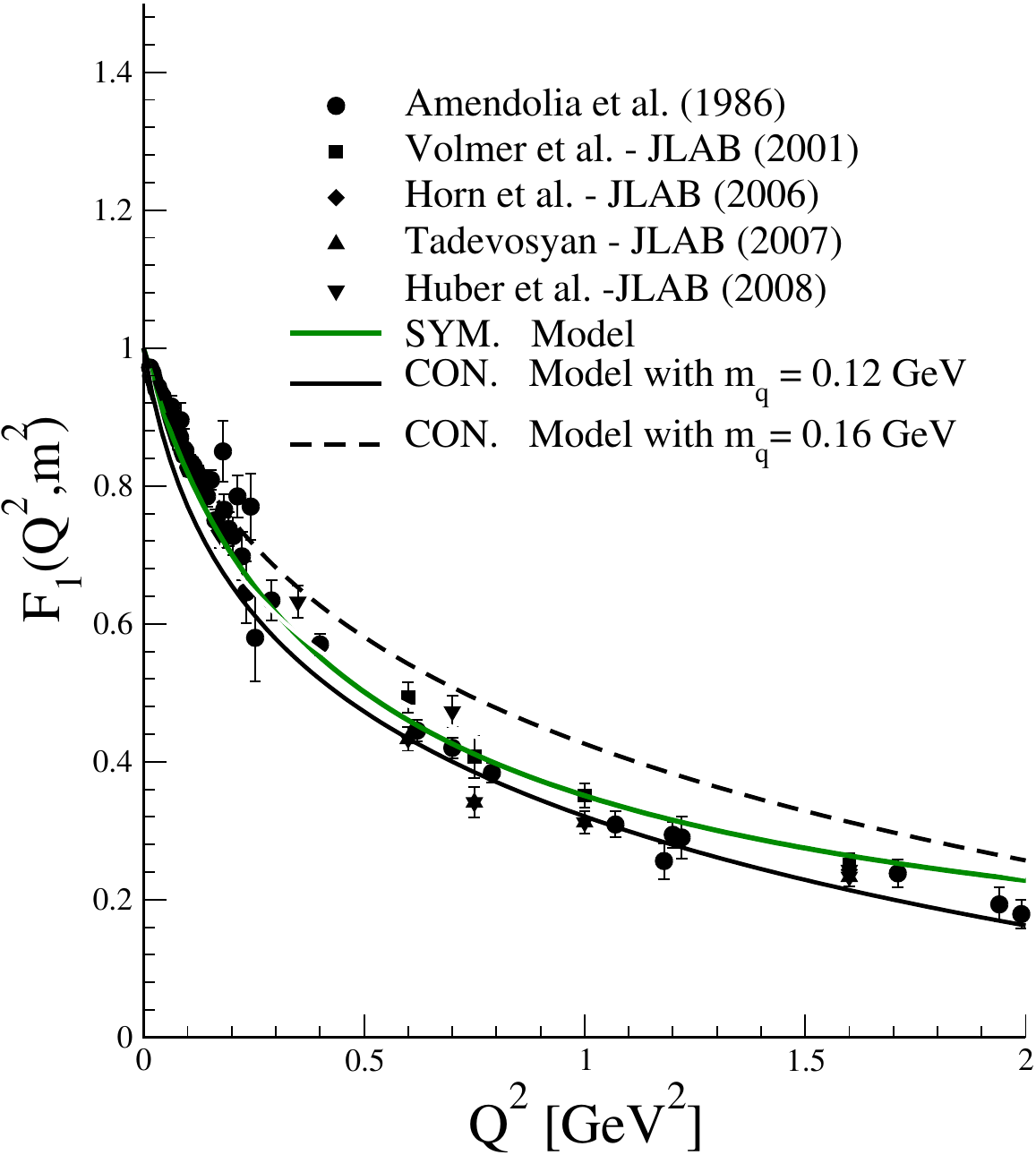,width=6.50cm,angle=0}   
 \caption{Pion electromagnetic on-mass shell form factor, $F_1(Q^2,m^2_\pi)$, 
 for the SYM and CON models, compared with experimental data.
 Black solid and dashed lines represent the CON model
 with $m_q=0.12\,$GeV ($g_{\pi q \bar{q}}=2.43$) and $m_q=0.16\,$~GeV ($g_{\pi q \bar{q}}=2.73$),
  respectively. Green line represents the SYM model.
 }  
	\label{Fig:f1onshell}
 \end{center} 
 \end{figure}
 
As a point of reference, we present the on-shell form factor 
	$F_1(Q^2,m^2_\pi)=F_\pi(Q^2)$  in Fig.~\ref{Fig:f1onshell} 
for both CON and SYM $\pi q{\bar q}$ vertices, namely,
$\Gamma_\pi (k, p) = (\Gamma^{\rm CON}_\pi, \Gamma^{\rm SYM}_\pi)$~\cite{deMelo1999,deMelo2002, Choi2019}, comparing the results with the experimental data. 
These calculations essentially capture the dispersion in
the data up to momentum transfers of 2\,GeV$^2$ within the model uncertainty.

\begin{figure}[htb]
	\begin{center}
		\epsfig{figure=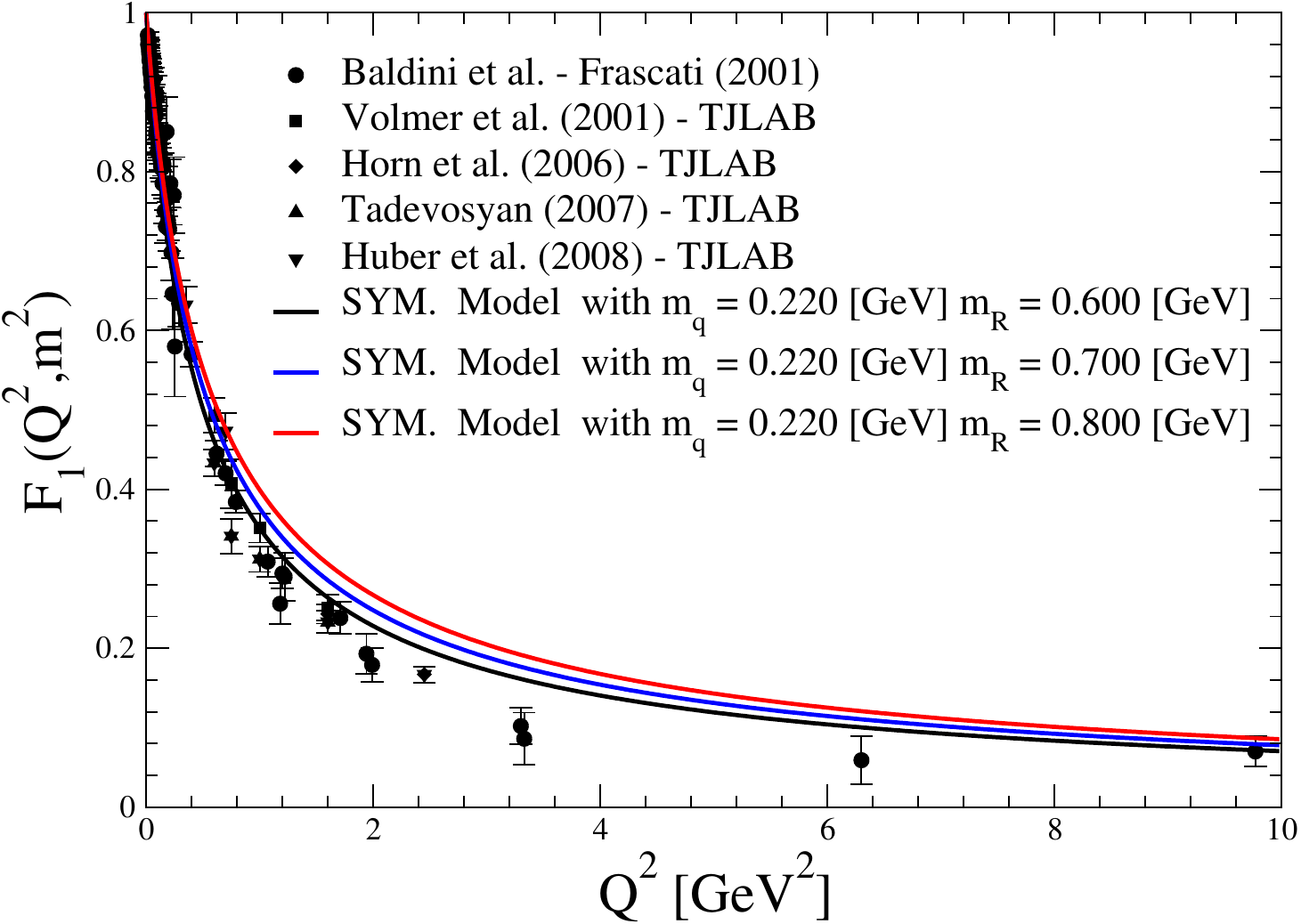,width=7cm,angle=0} 
		\caption{Pion electromagnetic form factor for the SYM model with regulator masses $m_R=$0.6, 0.7 and 0.8\,GeV for fixed constituent quark mass of 0.22\,GeV.}
		\label{fig:fpiSYMmR}
	\end{center} 
\end{figure}

	To illustrate the sensitivity of the SYM model to
	 the regularization mass in the pion form factor, we compare the results
	  for  $m_R$ values of 0.6, 0.7, and 0.8~GeV, with a fixed $m_q=0.220$ GeV, as shown in
	   Fig.~\ref{fig:fpiSYMmR}. The pion decay constant, $f_\pi$, varied by 
	   about 10\%, and the charge radius also changed by approximately 10\%, as presented in Table~\ref{tab:fpirpimR}. Notably,
	    $f_\pi$ is almost perfectly reproduced for $m_R=$0.6\,GeV, 
	    while the pion charge radius matches the experimental data for $m_R=$0.8\,GeV.
	     This variation produces a band 
	     that is consistent with the current experimental uncertainties. Interestingly,
	      the region of $1\lesssim Q^2\lesssim 5\,\text{Gev}^2$ 
	      shows the 
	      	highest sensitivity to changes in $m_R$, giving a variation 
	      resulting in a 15\%-20\% variation in the form factor, 
	      which then decreases outside this range.  This indicates that
	      the model is most sensitive to its parameters 
	      for momentum transfers around a few times the value of $m_R$.

\begin{table}[]
    \centering
\begin{tabular}{c|c|c|c|c}
	\hline\hline
$m_R$ [GeV] &  $f_\pi$ [MeV]  &  $\Delta f_\pi $(\%) &  $ r_\pi$[fm] 
& $\Delta r_\pi~(\%)$
    \\
	\hline
0.6	&  92.4     & 0.1  & 0.736   & 8.7        \\
0.7  &  97.0    &  4.9    &  0.695  &   3.4     \\
0.8  &   100.9   &  8.5    & 0.675   &   0.4   \\
	\hline\hline
\end{tabular}
    \caption{ Results for the pion decay constant, $f_\pi$, and charge radius, $r_\pi$,   for $m_q=0.22\,$GeV and different values of the regulator mass $m_R$. 
    	The fractional percent deviations are  
    $\Delta f_\pi =   \left| \frac{f^{SYM}_\pi  -  f^{\rm Exp.}_\pi  }{  f^{SYM}_\pi} \right|  \times 100 $ and 
    $\Delta r_\pi = \left| \frac{r^{SYM}_\pi  -  r^{\rm Exp.}_\pi  }{  r^{SYM}_\pi}  \right| \times 100 $.}
    \label{tab:fpirpimR}
\end{table}

\subsection{Off-shell form factors}
\label{subsec:offshelff}

\begin{figure}
\begin{center}
\includegraphics[height=6cm,width=6cm]{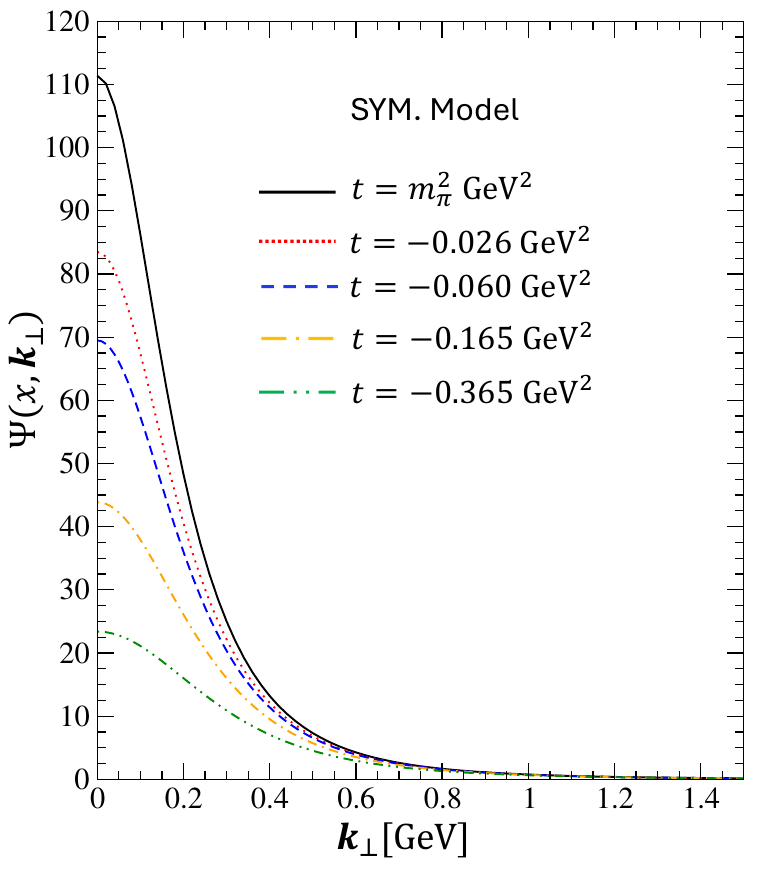}
\caption{\label{wf2dkpe} Pion wave function $\Psi(x, k_\perp)$
	 obtained from the SYM model, evaluated at fixed values of 
$x = 0.5$ for various values of $t$ as a function of $k_\perp$.}
\end{center}
\end{figure}

\begin{figure}
\begin{center}
\includegraphics[height=6cm,width=6cm]{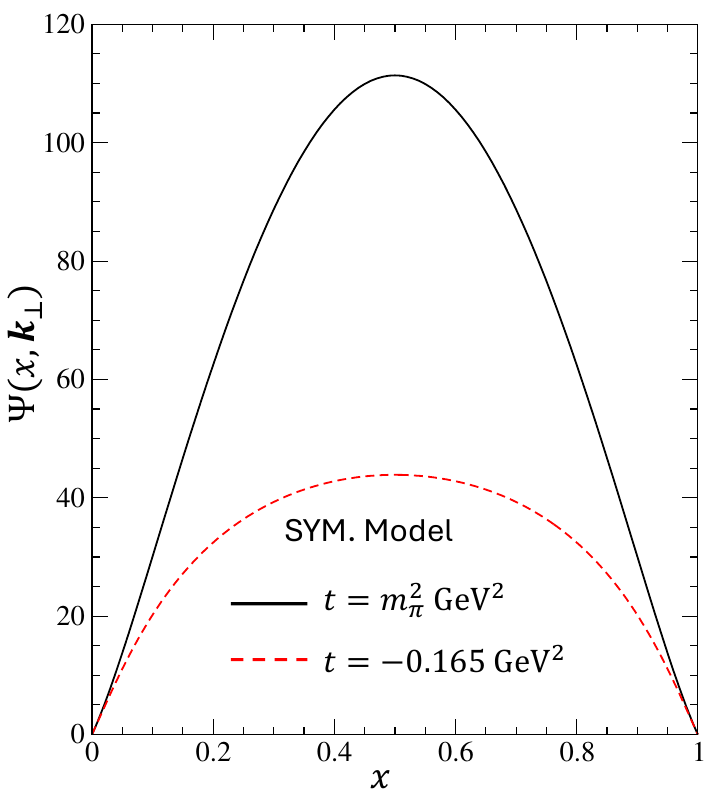}
\caption{\label{wf2dxx} Pion wave function $\Psi(x, k_\perp)$ obtained
	 from the SYM model, evaluated at fixed values of 
$k_\perp=0$ for the two values of $t=(m^2_\pi, -0.165)$ 
GeV$^2$ as a function of $x$.}
\end{center}
\end{figure}

\begin{figure}
\begin{center}
\includegraphics[height=5cm, width=8cm]{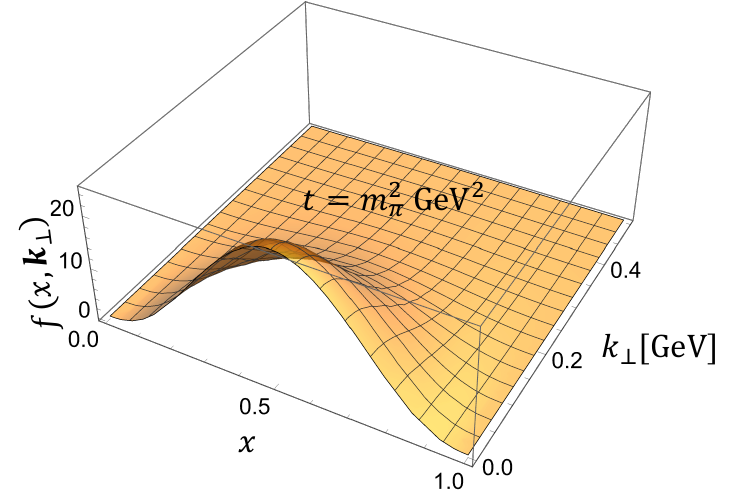}
\includegraphics[height=5cm, width=8cm]{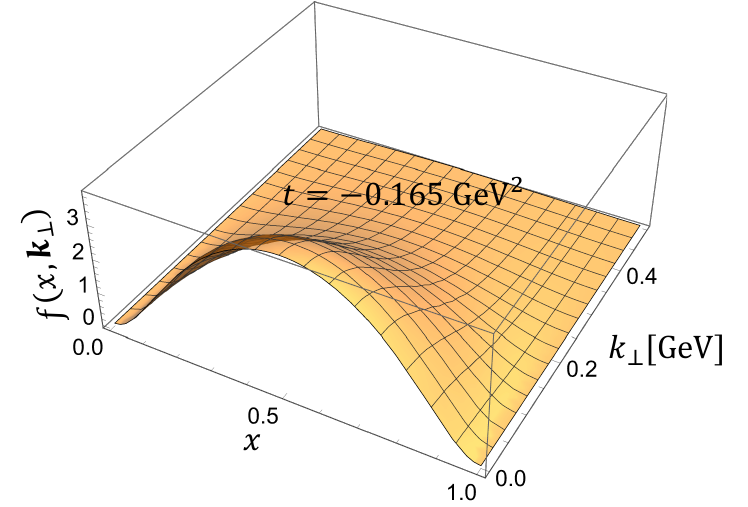}
\caption{\label{3dwavefunction} Twist-2 pion TMD $f(x, k_\perp)$
	 obtained from the SYM model, evaluated at a fixed value of 
$Q^2=0$,  for the two values of $t=m^2_\pi$ (top) and -0.165 GeV$^2$ (bottom).}
\end{center}
\end{figure}

\begin{figure}
\begin{center}
\includegraphics[height=6cm, width=6cm]{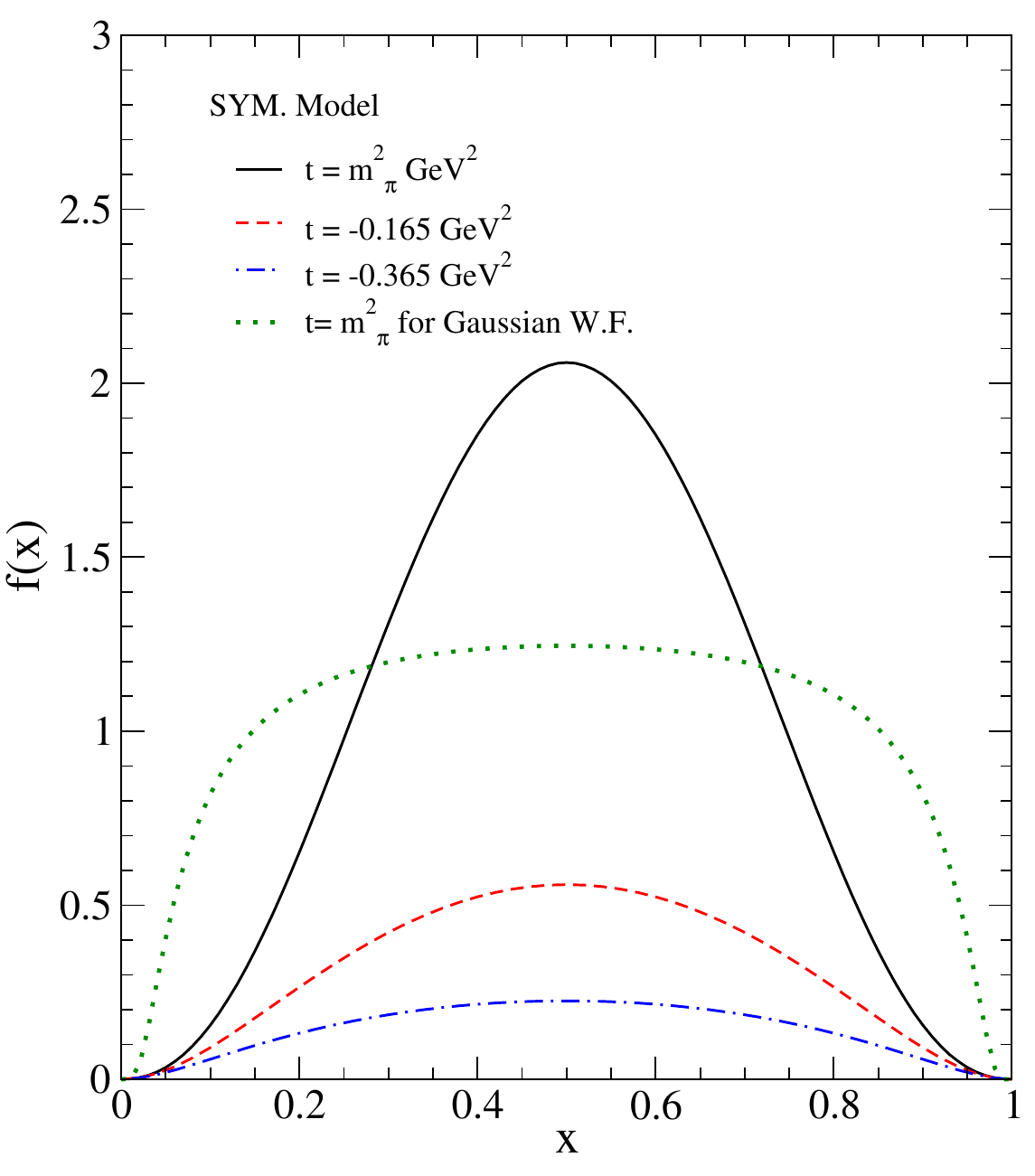}
\caption{\label{Fig:pantipargamma} Twist-2 pion PDF
	 $f(x)$ obtained from the SYM model, evaluated at a fixed value of 
$Q^2=0$,  for various values of $t$.  For comparison,
 the light-front quark model result for the on-shell pion, obtained
  from the Gaussian wave function~\cite{choi24},
is also included.}
\end{center}
\end{figure}

Figures~\ref{wf2dkpe} and~\ref{wf2dxx} 
show the pion wave function $\Psi(x, k_\perp, t= p^+p^- -p_\perp^2)$
 obtained from the SYM model, evaluated for various $t$ values.
In particular, Figure~\ref{wf2dkpe} displays the wave function at fixed $x=0.5$ as a function of
 $k_\perp$ for $t=(m^2_\pi \approx 0.020, -0.026, -0.060, -0.165, -0.365)$ GeV$^2$, while Figure~\ref{wf2dxx} presents it at fixed $k_\perp=0$ for the 
two values $t=(m^2_\pi, -0.165)$ GeV$^2$ as a function of $x$.
As one can see from both figures, the off-shell wave function decreases as $t$ deviates 
from the on-mass shell ($t=m^2_\pi$) for both the $k_\perp$ and $x$ variables.

In Fig.~\ref{3dwavefunction}, we show the unpolarized twist-2  pion TMD $f(x, k_\perp)$
 obtained from the SYM model 
for the two values: $t=m^2_\pi$ (top panel) and -0.165 GeV$^2$ (bottom panel).
For the  twist-2 TMD $f(x, k_\perp)$, the distribution of a quark with a longitudinal momentum fraction $x$ is 
identical to the distribution of an antiquark with a longitudinal momentum fraction
 $1-x$,  i.e. $f^q(x, k^2_\perp)=f^{\bar q}(1-x, k^2_\perp)$. 
Moreover, the pion TMDs for quark and antiquark are the same, 
resulting in a momentum distribution that is symmetric with respect to $x=1/2$, as expected for the 
bound state of an equal mass quark-antiquark system.

We also display in Fig.~\ref{Fig:pantipargamma} the twist-2 pion 
PDF $f(x)$ obtained from the SYM model for a few different values of $t$. 
For the comparison with the more elaborated light-front quark 
	model (LFQM) based on the variational analysis of the  
meson mass spectra, we also display the result obtained from the Gaussian 
trial wave function for the on-shell pion~\cite{choi24}. 
We note that the pion PDFs obtained from the SYM model wave function and from the Gaussian model wave function determined from the variational analysis of meson mass spectra~\cite{choi24} are quite different indicating the sophistication of the more phenomenologically successful 
LFQM with respect to the simple SYM model presented in this work. 
The essential characteristic of flattening the PDF for the central regions of $x$ and enhancing the end point regions of $x$
in the Gaussian radial wave function used in~\cite{choi24} is due to the inclusion 
of the Jacobi factor (see Eq. (8) in~\cite{choi24}),
 which is crucial for the rotational invariance of the self-consistent LFQM wave
  function as discussed in Refs.~\cite{Jafar1,Jafar2}. 
While this  calls for a more elaborated LFQM analysis, we limit ourselves in this  work  to exploring the 
essential characteristic of the off-shell light-front wave functions 
using a rather simple SYM model.

\begin{figure}[t]
\begin{center}
\epsfig{figure=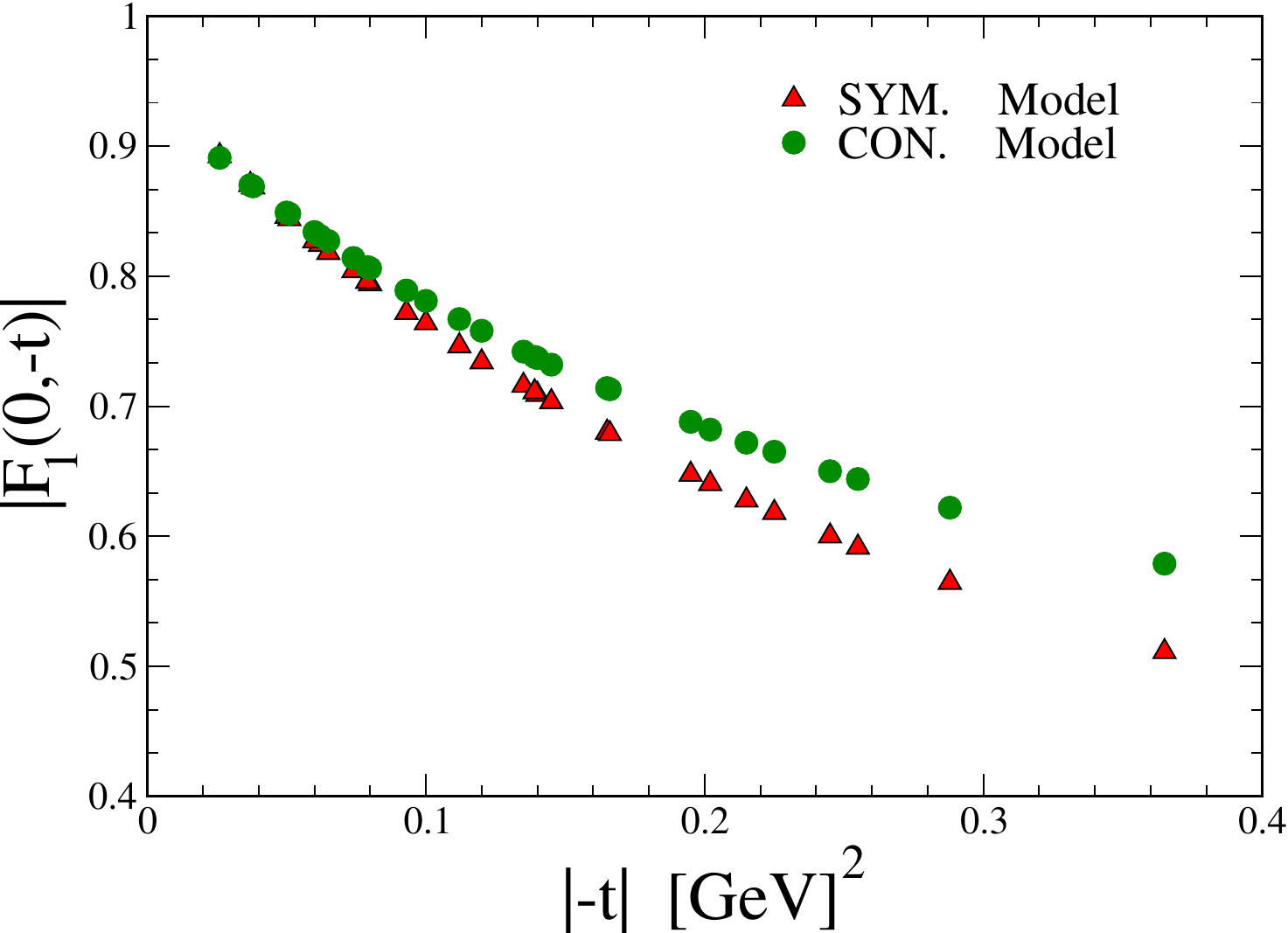,width=6.89cm,angle=0}

\vspace{0.85cm}

\epsfig{figure=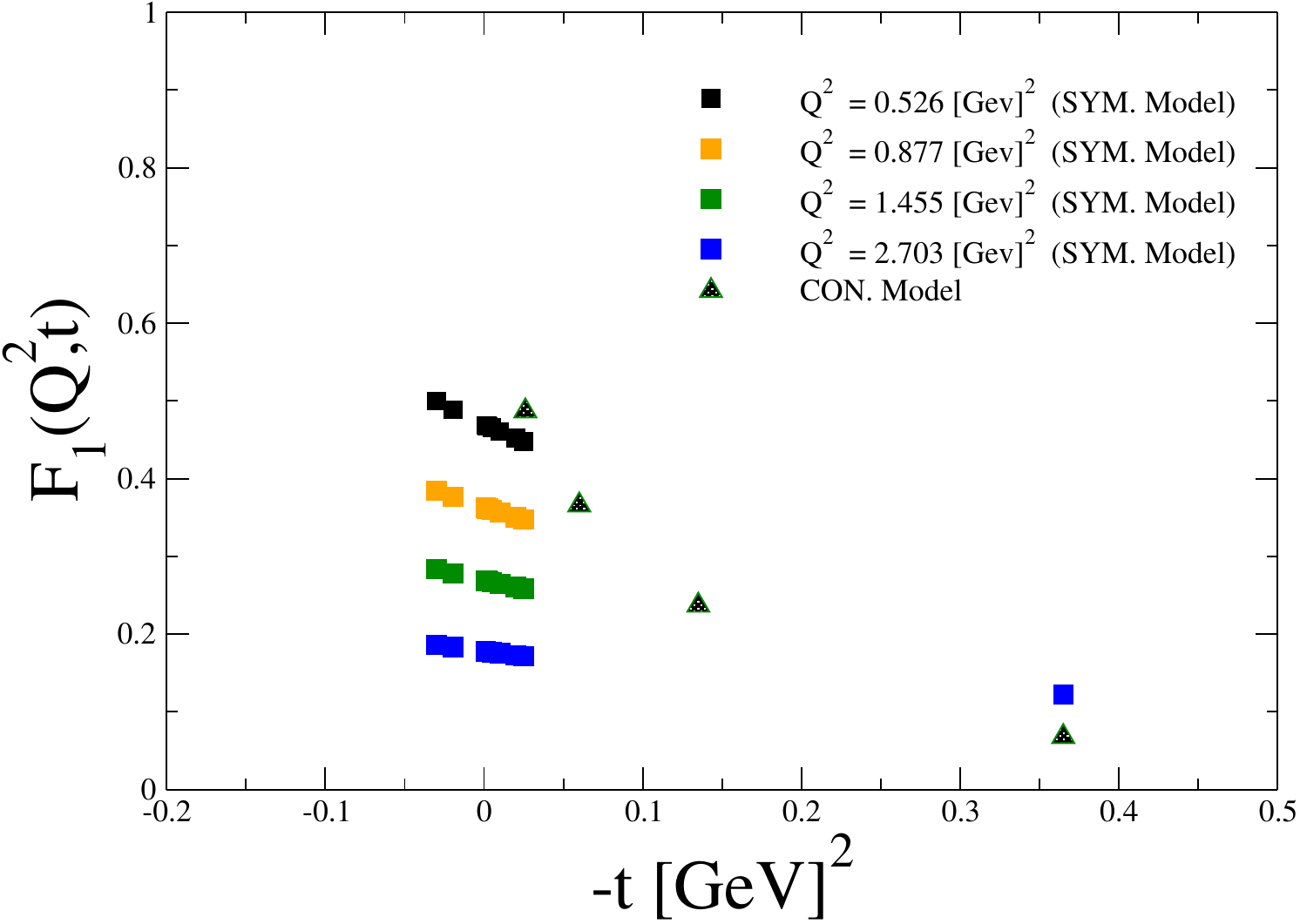,width=6.89cm,angle=0}

\caption{
Pion off-shell EM form factor $F_1(Q^2,t)$ 
obtained from the CON~\cite{Choi2019} and SYM
 models~\cite{deMelo2002} plotted as a function of $-t$. 
The top panel shows the result of $F_1(0,t)$, and  the bottom
 panel shows $F_1(Q^2,t)$ for selected values of $Q^2=0.526\,,0.877\, ,1.455$ 
 and 2.703$\,[\text{GeV/c}]^2$.
}
\label{Fig:F1th}
\end{center} 
\end{figure}

We shall now present our results from the study of the pion on-shell
 EM form factors using the microscopic constituent quark models 
based on the symmetric $\pi q{\bar q}$ vertex,
	 $\Gamma^{\rm SYM}_\pi$ given
	  by  Eq.~\eqref{eq:sym}.
We will compare it with the constant vertex
 $\Gamma^{\rm CON}_\pi$ given in Eq.~\eqref{eq:cov}.  
The computed $F_1(0,t)$ and $F_1(Q^2,t)$ for the two models are tabulated in Table~\ref{table1} and illustrated in Fig.~\ref{Fig:F1th}. 
In the top panel of Fig.~\ref{Fig:F1th},
we show $|F_1(0, -t)|$ as a function 
of $|-t|$ for the models $(\Gamma^{\rm CON}_\pi,\Gamma^{\rm SYM}_\pi)$. It is noted that
$F_1(0,0) <1$ since the models are normalized such that $F_1(0,m^2_\pi)=1$. 

The overall model dependence appears at the level of a few percent.
Both the constant and symmetric vertex models show good agreement for $-t<0.1\,$GeV$^2$.
In the bottom panel  of Fig.~\ref{Fig:F1th}, $F_1(Q^2,t)$  is plotted as a function of $t$ for
selected values of $Q^2$, such as $0.526\,,0.877\, ,1.455$ and 2.703\,[GeV/c]$^2$,  taken from Table~\ref{table1}.
We observe a mild model  dependence that 
becomes more pronounced as the momentum transfer increases. 
It is important to note 
that $F_1(0,t)$ is essential for extracting the off-shell form factor 
$g(Q^2,t)$ from the experimental value of $F_1(Q^2,t)$, 
thereby introducing some model dependence into this extraction.

 \begin{figure}[t]
 	\begin{center}
 
\vspace{0.5cm} 
 	
\epsfig{figure=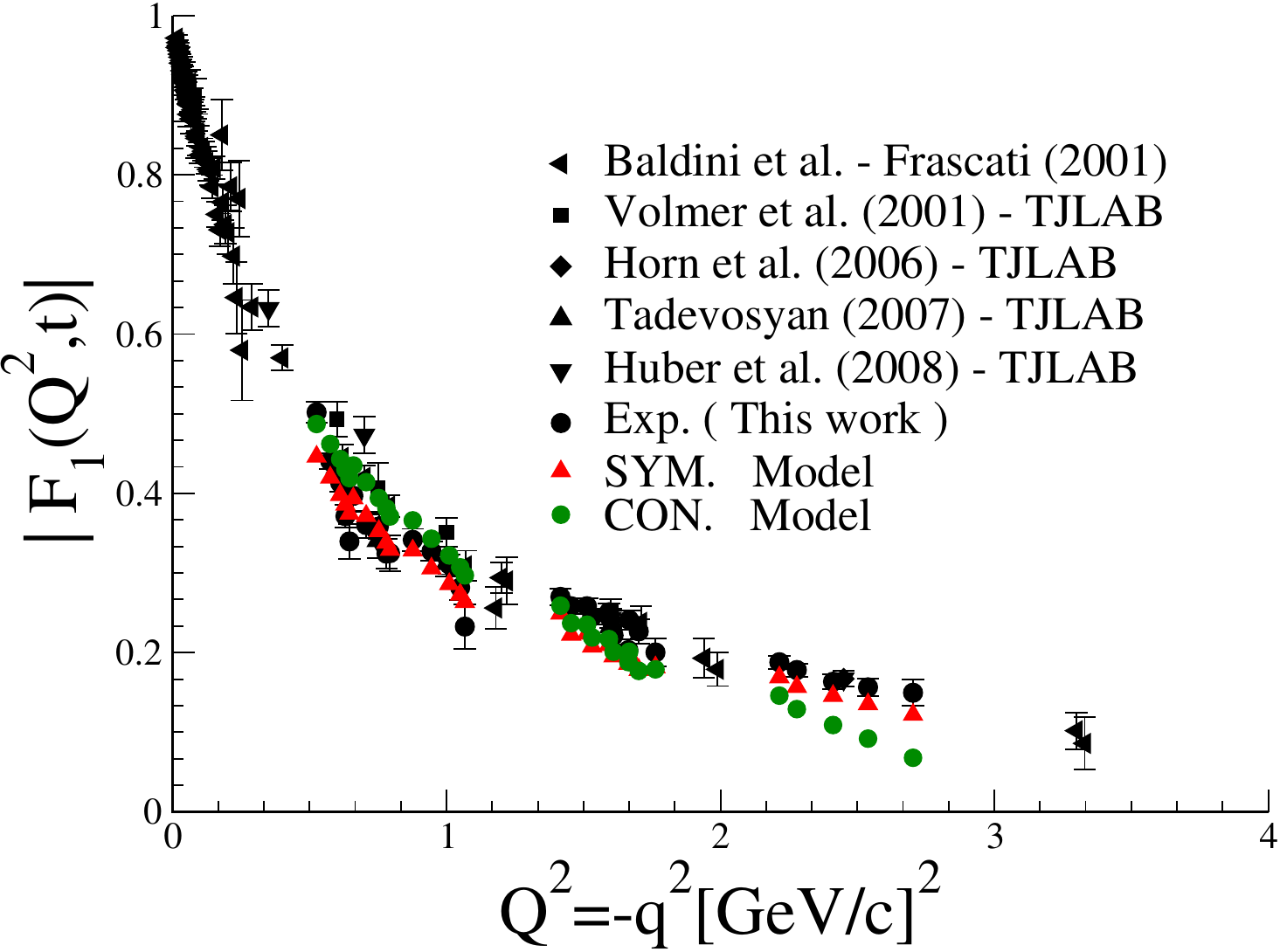,width=6.6cm,angle=0}

\vspace{0.8cm}

\epsfig{figure=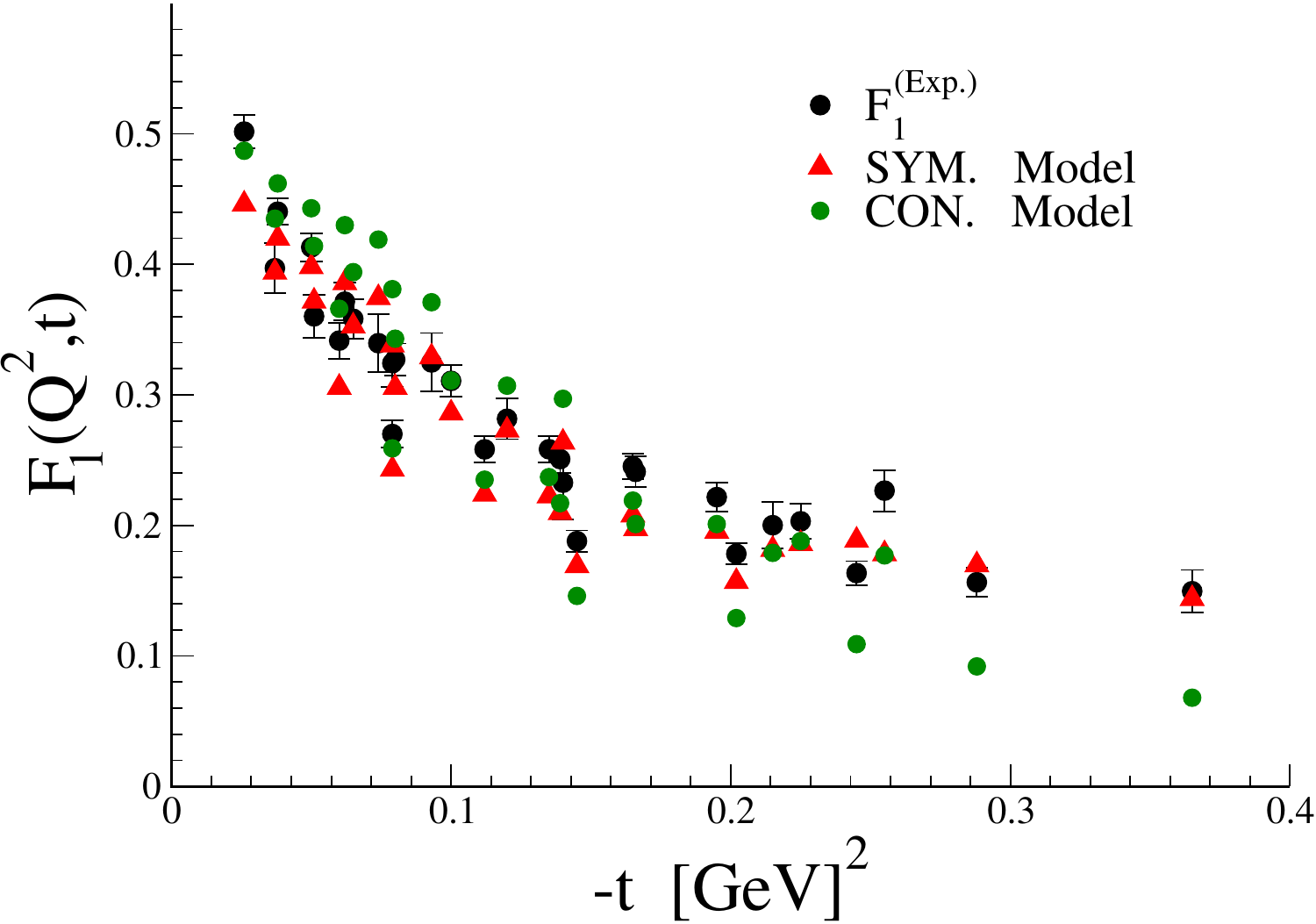,width=7cm,angle=0} 
 \caption{Pion off-shell EM form factor $F_1(Q^2,t)$.
 Top panel: $F_1(Q^2,t)$ as a function of $Q^2$ extracted from
  the experimental cross sections~\cite{Blok2008}, 
   CON, and SYM models.
Bottom panel: $F_1(Q^2,t)$ as a function of $-t$ for the same models.
The values of $(Q^2,t)$ are from Table~\ref{table1} 
in Appendix~\ref{app:tables}, along with the 
experimental and theoretical results from Table~\ref{table3}.
}
 \label{f1exp}
 \end{center} 
 \end{figure}

 \begin{figure}[t]
    \begin{center}
\vspace{0.5cm}   	
\epsfig{figure=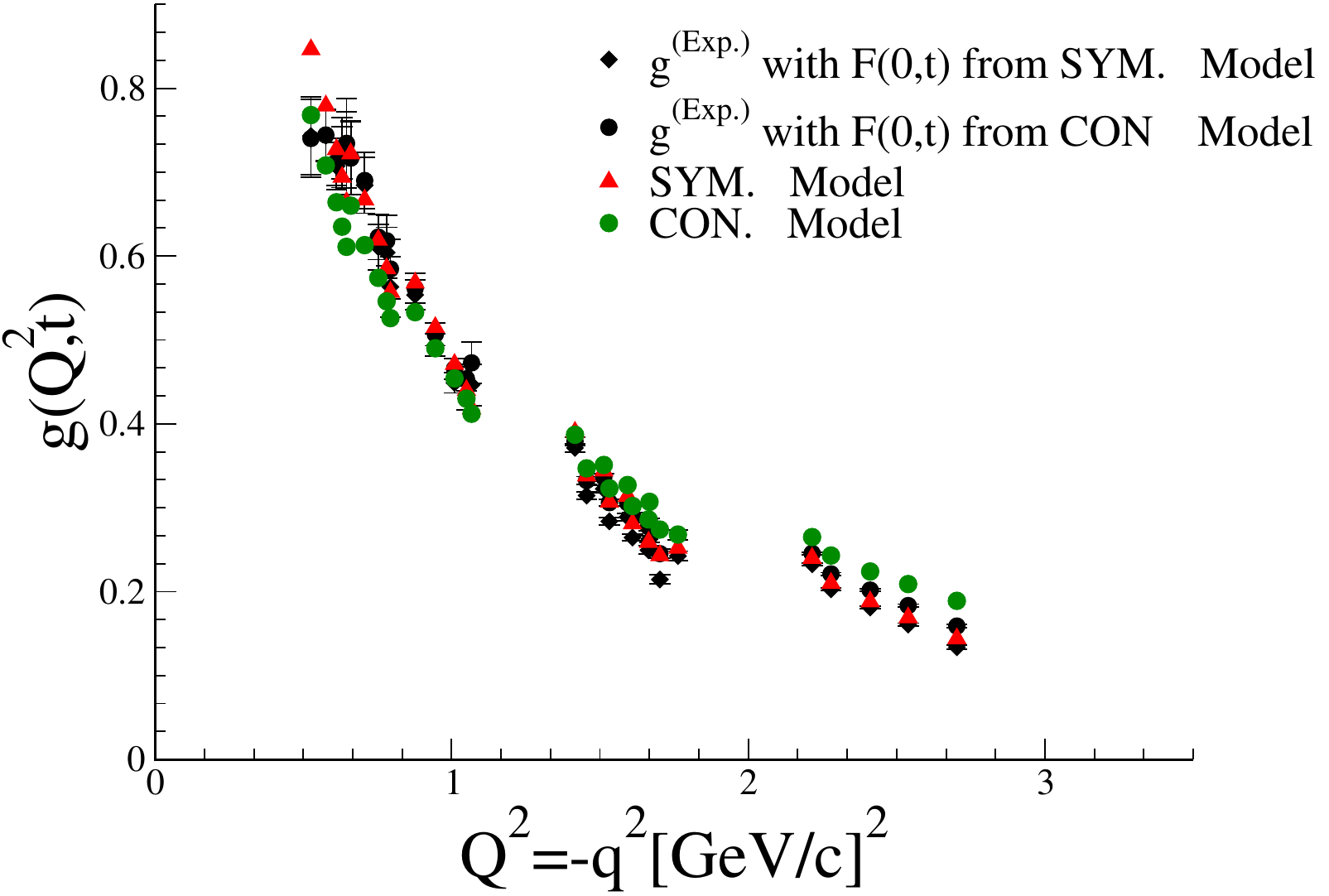,width=7.cm,angle=0}
\\ \vspace{1.002cm}
\epsfig{figure=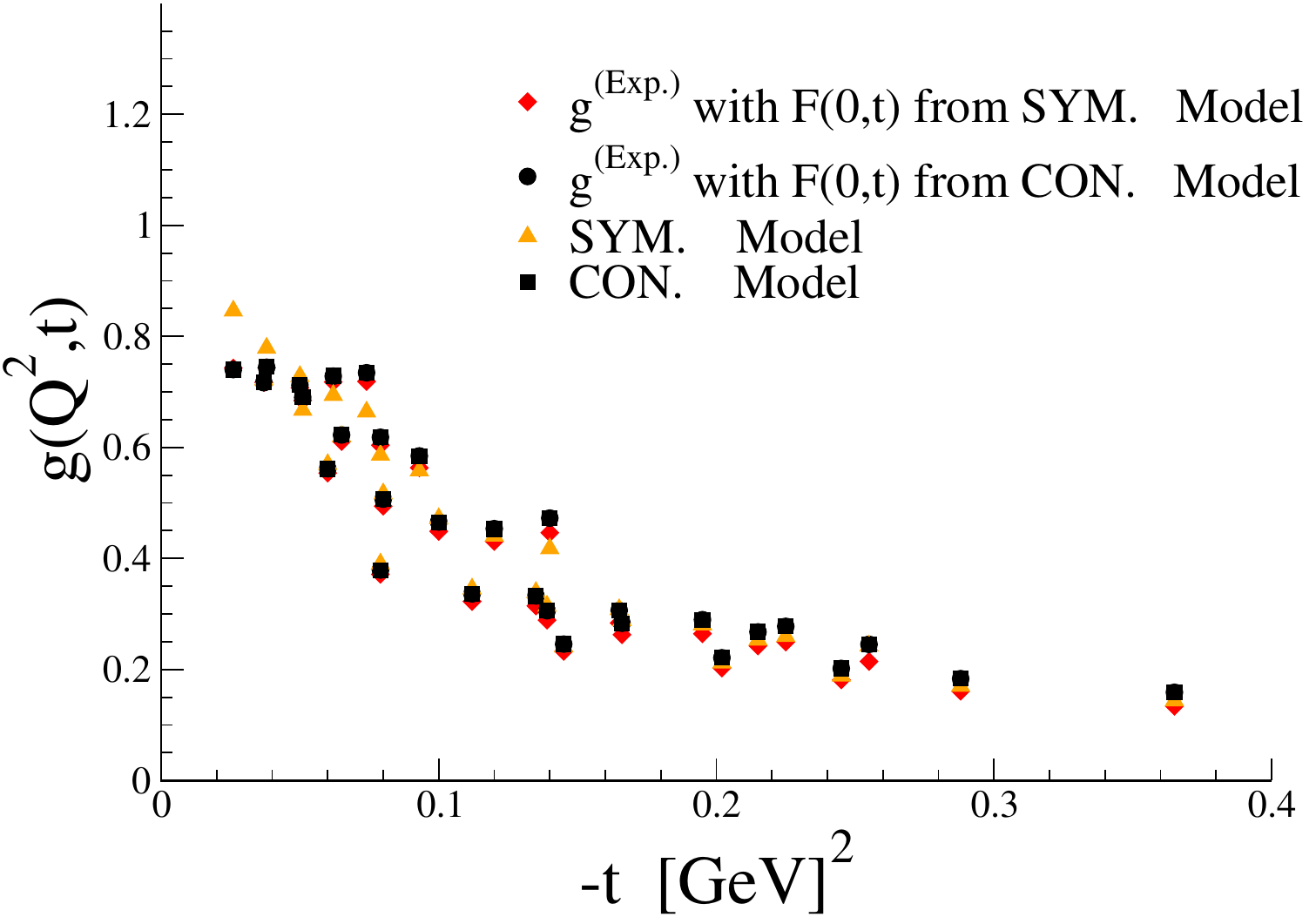,width=7.cm,angle=0} 
 \caption{
 Pion off-shell EM form factor $g(Q^2,t)$.
 Top panel: $g(Q^2,t)$ as a function of $Q^2$, showing the extracted result from the
experimental data~\cite{Blok2008} with the theoretical inputs
 $F_1(0,t)$ from the  CON and SYM models, as well as the theoretical results from these two models. 
Bottom panel: $g(Q^2,t)$ as a function of $-t$ for the same models.
The values of $(Q^2,t)$ are from Table~\ref{table2} 
in Appendix~\ref{app:tables}, along with the extracted
experimental and theoretical results.
}
 \label{Fig:gexp1}
 \end{center} 
 \end{figure}

In Fig.~\ref{f1exp}, we display the pion off-shell EM form factor $F_1(Q^2,t)$ 
obtained from the two different models. The top panel shows
 $F_1(Q^2,t)$ as a function of $Q^2$ for all values of $t$,
and the bottom panel shows it as a function of $-t$.  These results are
 compared with experimental values extracted from the cross section in Ref.~\cite{Blok2008} 
and summarized in Ref.~\cite{Choi2019}, as well as in Table~\ref{table1}.
We  should note that the extracted  experimental values of $F_1(Q^2,t)$  
are obtained from the dominant pion pole in the Chew-Low approach 
to the photo-production amplitude in the exclusive version of the Sullivan process, 
as discussed in Sec.~\ref{subsec:xsection}.

The symmetric  vertex model exhibits smaller dispersion in relation to the extracted  experimental values. 
This is evident in the upper panel  of Fig.~\ref{f1exp} for $Q^2\gtrsim 2\,$GeV$^2$ and also in the lower panel, 
where the spread is further influenced by changes in $Q^2$. 
 Overall, the two models for the off-shell form factor, $F_1(Q^2,t)$, 
cover the dispersion of the extracted experimental values similarly 
to the on-shell form factor, $F_1(Q^2, m^2_\pi)$,  as shown in Fig.~\ref{Fig:f1onshell}.

 \begin{figure*}[t]
 	\begin{center}
\epsfig{figure=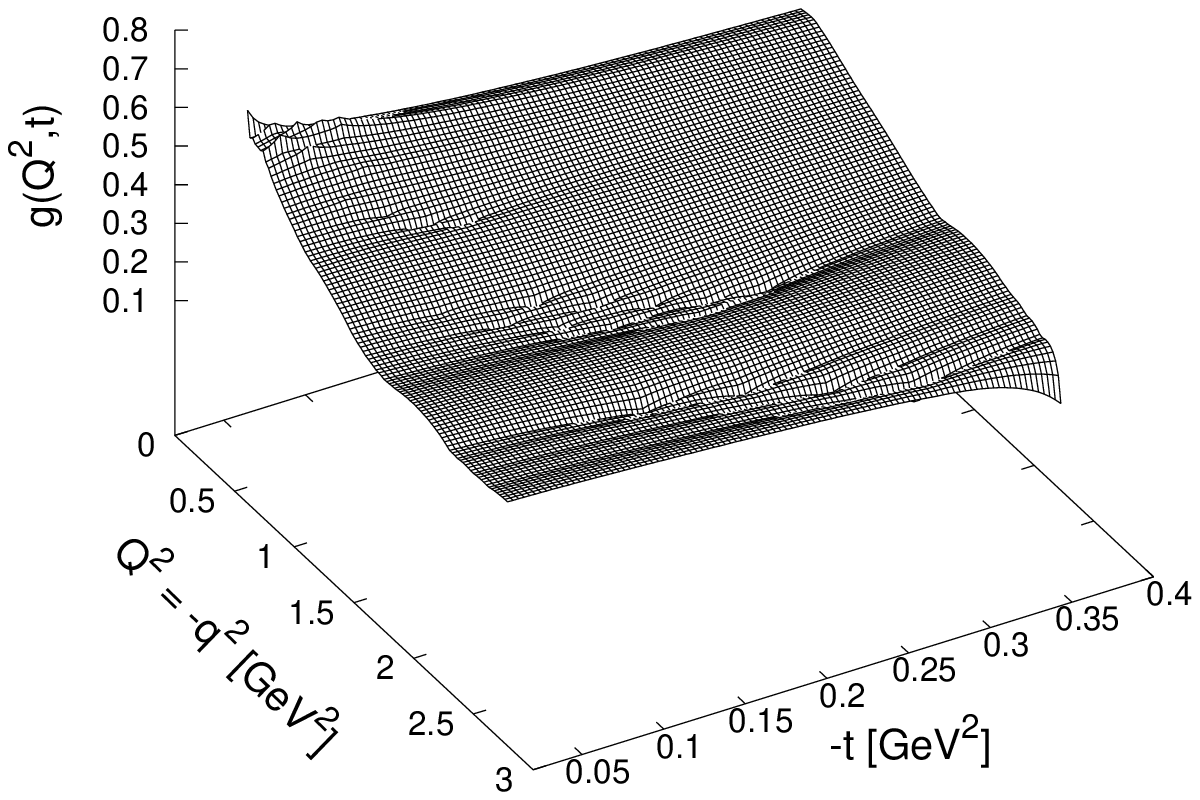,width=6.1cm,angle=-90}
\epsfig{figure=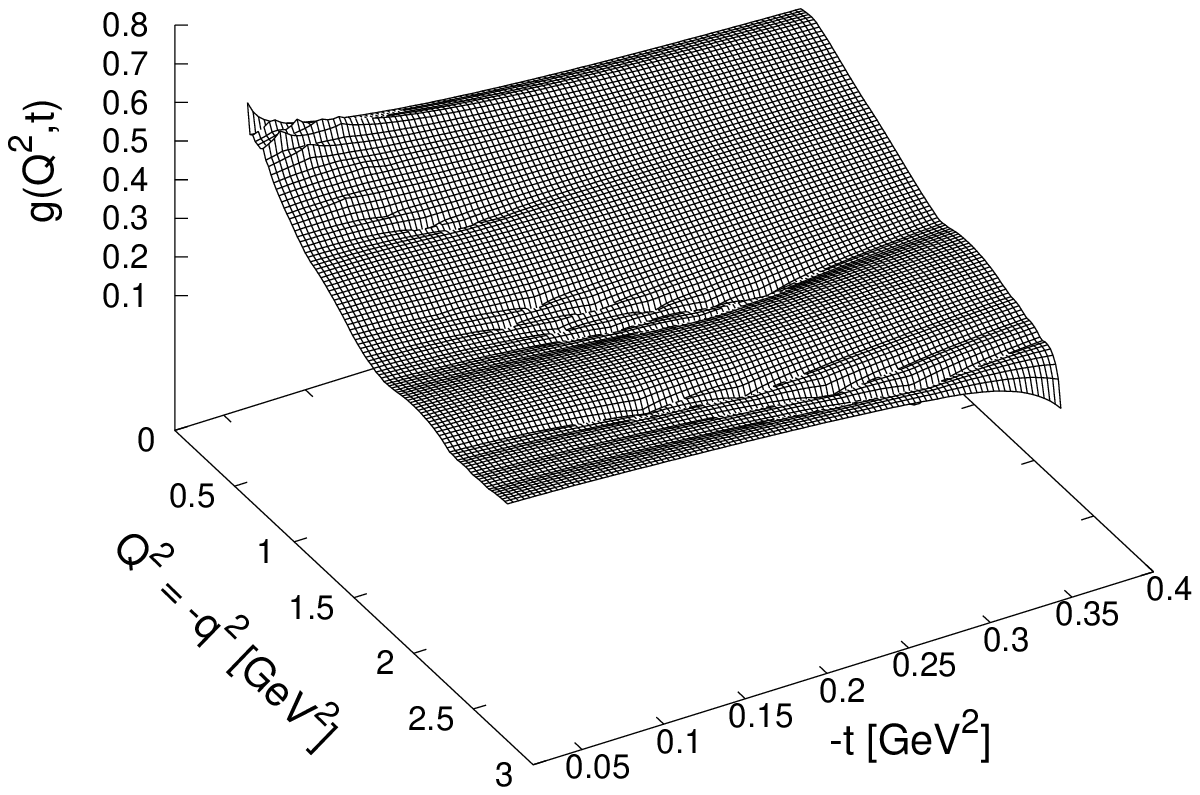,width=6.1cm,angle=-90} 
   \caption{
The 3D plots of the form factor $g(Q^2,t)$ extracted from the
experimental data~\cite{Blok2008}
with the theoretical  inputs $F_1(0,t)$ from the CON (left panel) 
and SYM (right panel) models. The extracted values  are given in Table~\ref{table2} 
from Appendix~\ref{app:tables} and labeled as 
$g^{\rm CON}(Q^2,t)$
and
$g^{\rm SYM}(Q^2,t)$, respectively.
} 
 \label{Fig:gexp3d}
\end{center} 
\end{figure*}

In Fig.~\ref{Fig:gexp1}, we present the results for the form factor
 $g(Q^2,t)=(g^{\rm CON}, g^{\rm SYM})$ obtained 
from the $\Gamma_\pi=(\Gamma^{\rm CON}_\pi, \Gamma^{\rm SYM}_\pi)$
vertices. Additionally,  $g^{\rm Exp}(Q^2,t)$, extracted from the experimental 
cross-sections listed in Table VII of Ref.~\cite{Blok2008}, is shown.
These are presented
as a function of $Q^2$ in the upper panel and $-t$ in the lower panel.
The extracted values for $g^{\rm CON}(Q^2,t)$, 
and $g^{\rm SYM}(Q^2,t)$
are also summarized in Table~\ref{table2} of Appendix~\ref{app:tables}.

 We should note that extracting $g(Q^2,t)$ requires knowledge 
 of $F_1(0,t)$, and the values of $F_1(0,t)$ are obtained from the two
 vertices presented in this work.  Given this, our extraction
 of $g(Q^2,t)$ from $F_1^{\rm Exp}(Q^2,t)$ inherently reflects some model dependence.
In the upper panel of Fig.~\ref{Fig:gexp1},
one can observe an approximate linear behavior  of $g(Q^2,t)$ 
at small $Q^2$ values, followed by a sharp increase as $Q^2$ approaches zero.
This behavior occurs because $g(Q^2,t)$ reaches the experimental value of $g^{\rm Exp}(0, m^2_\pi)=\la r^2_\pi\ra/6 =1.93(5)$,
 as shown in Table~\ref{table0}.
In the lower panel  of Fig.~\ref{Fig:gexp1}, the dependence of $g(Q^2,t)$ on $t$ is
 explored.  The dispersion observed in this plot for $g(Q^2,t)$
 reflects the  model dependence  arising from the different values of $F_1^{\rm Exp}(Q^2,t)$ 
associated with the model vertices $\Gamma_\pi$.
Despite  the model dependence of $g(Q^2,t)$ on $t$, the sharp increase in $g(Q^2,t)$
 as $Q^2\to0$ and $(-t)\to m^2_\pi$ is a common behavior across all the 
$\pi q{\bar q}$ vertices used in this work.
Quantitatively, the differences among the three different models in extracting
 $g(Q^2,t)$ are relatively mild ($\lesssim 7\%$) in the low $Q^2$ region but
increase steadily up to $10\%$ in the $2\leq Q^2\leq 3\,$GeV$^2$ range.

In Fig.~\ref{Fig:gexp3d}, we show the  3D plot of the extracted values 
of  $g(Q^2,t)$ for $g^{\rm CON}$ (left panel), 
and $g^{\rm SYM}$ (right panel),
illustrating their behavior with respect to $Q^2$ and $(-t)$.
The sharp behavior near $Q^2=t=0$ is consistent in both models, 
as previously noted in the discussion  of Fig.~\ref{Fig:gexp1}.  
The wiggles observed in the plot arise from the dispersion in 
$F_1^{\rm Exp}(Q^2,t)$, which is
 extracted from the longitudinal cross-section $d\sigma_{\rm L}/dt$~\cite{Blok2008}. 

 \begin{figure*}[t]
 	\begin{center}
\epsfig{figure=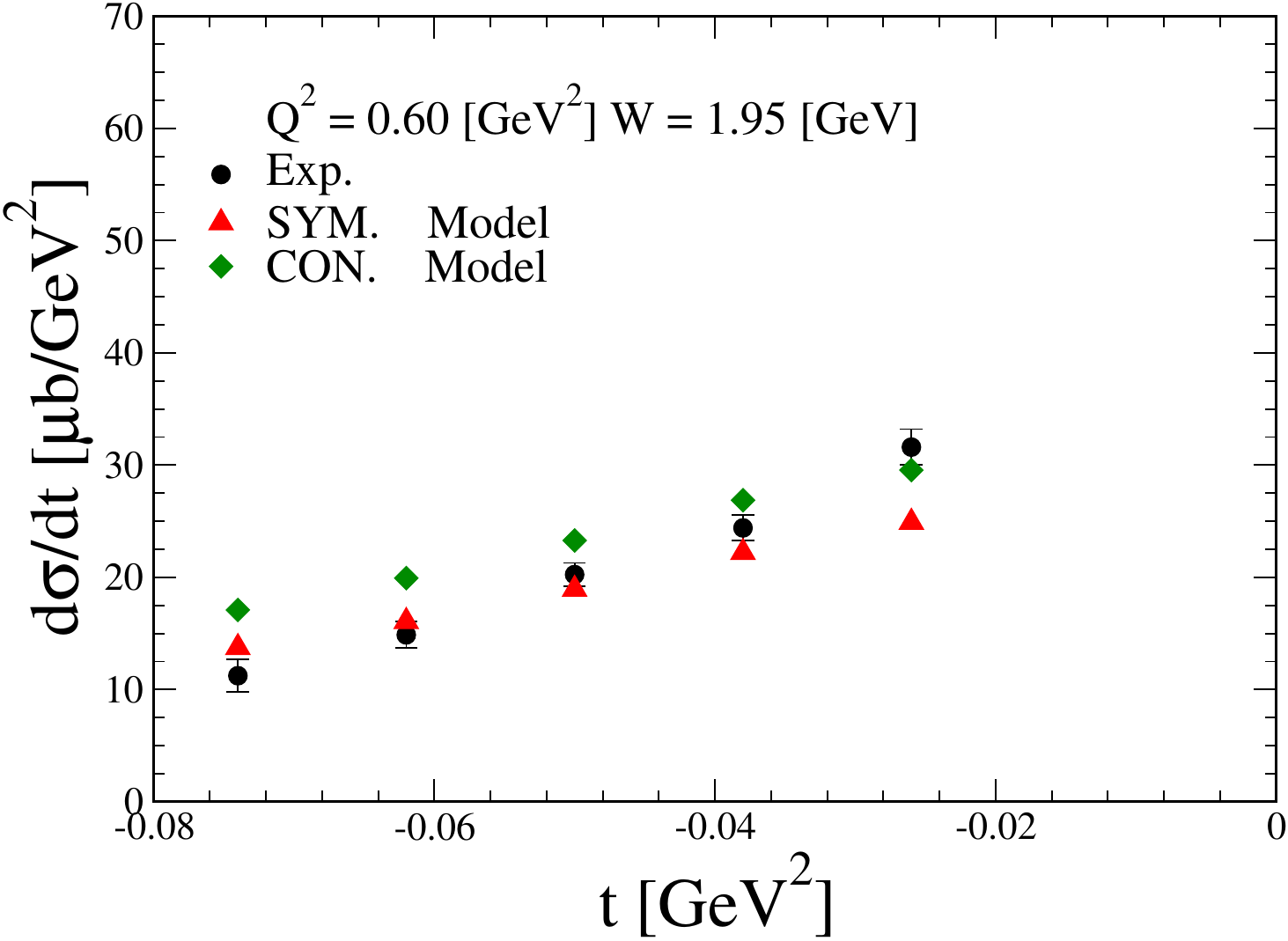,width=5.cm,angle=0}
\vspace{0.28cm}
\epsfig{figure=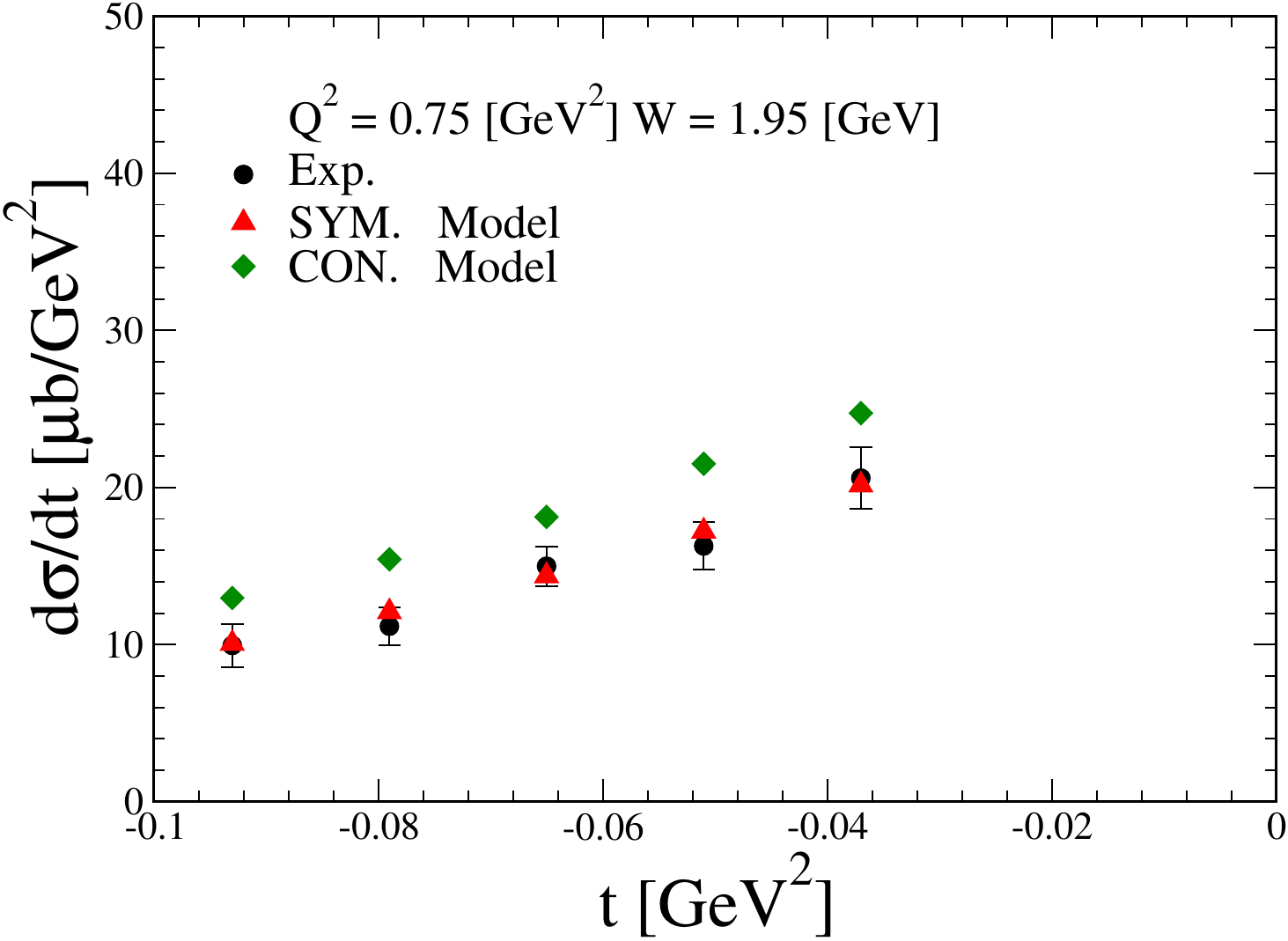,width=5.cm,angle=0} 
\vspace{0.28cm}
\epsfig{figure=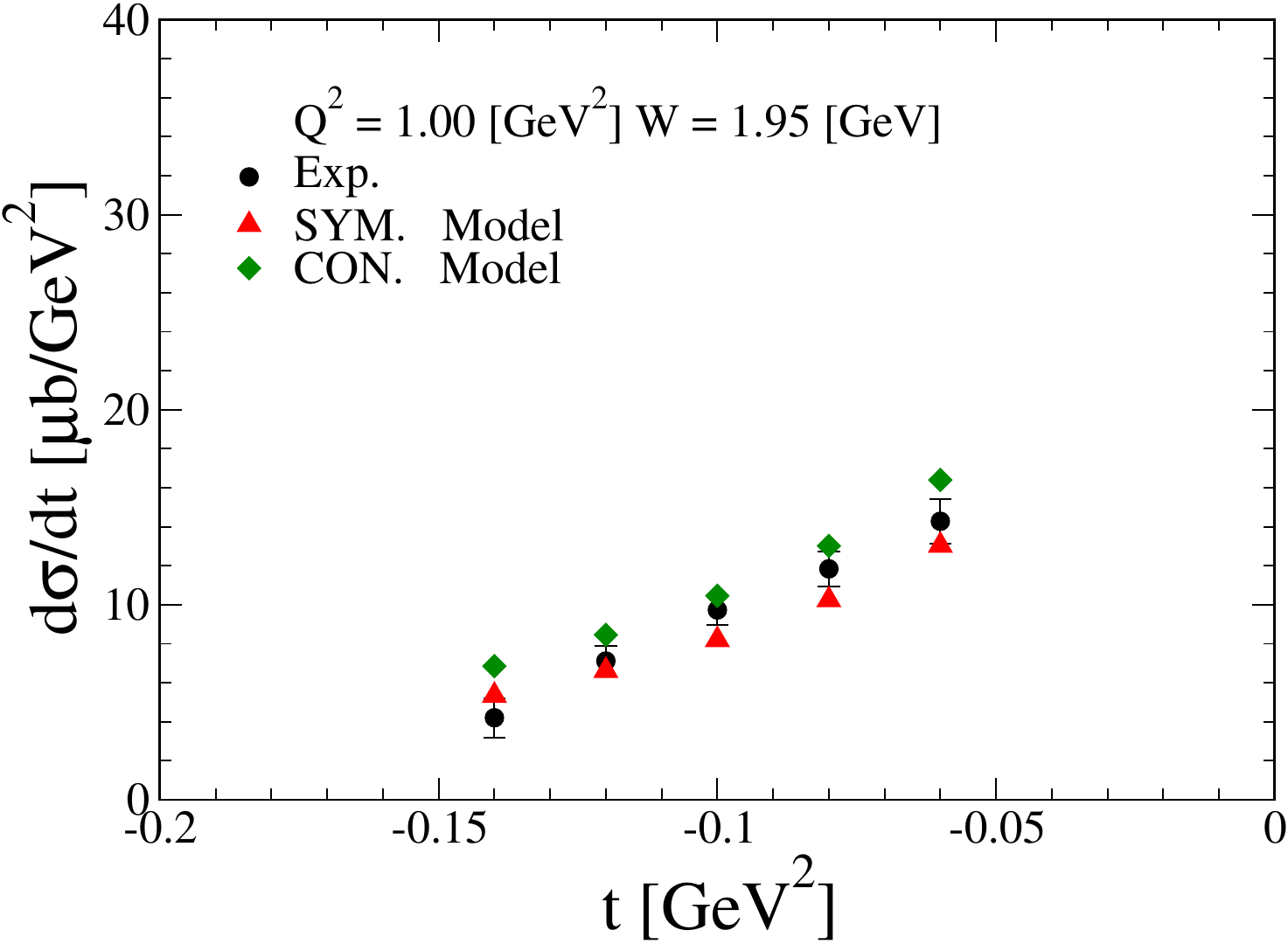,width=5.cm,angle=0}
\vspace{0.28cm}
\epsfig{figure=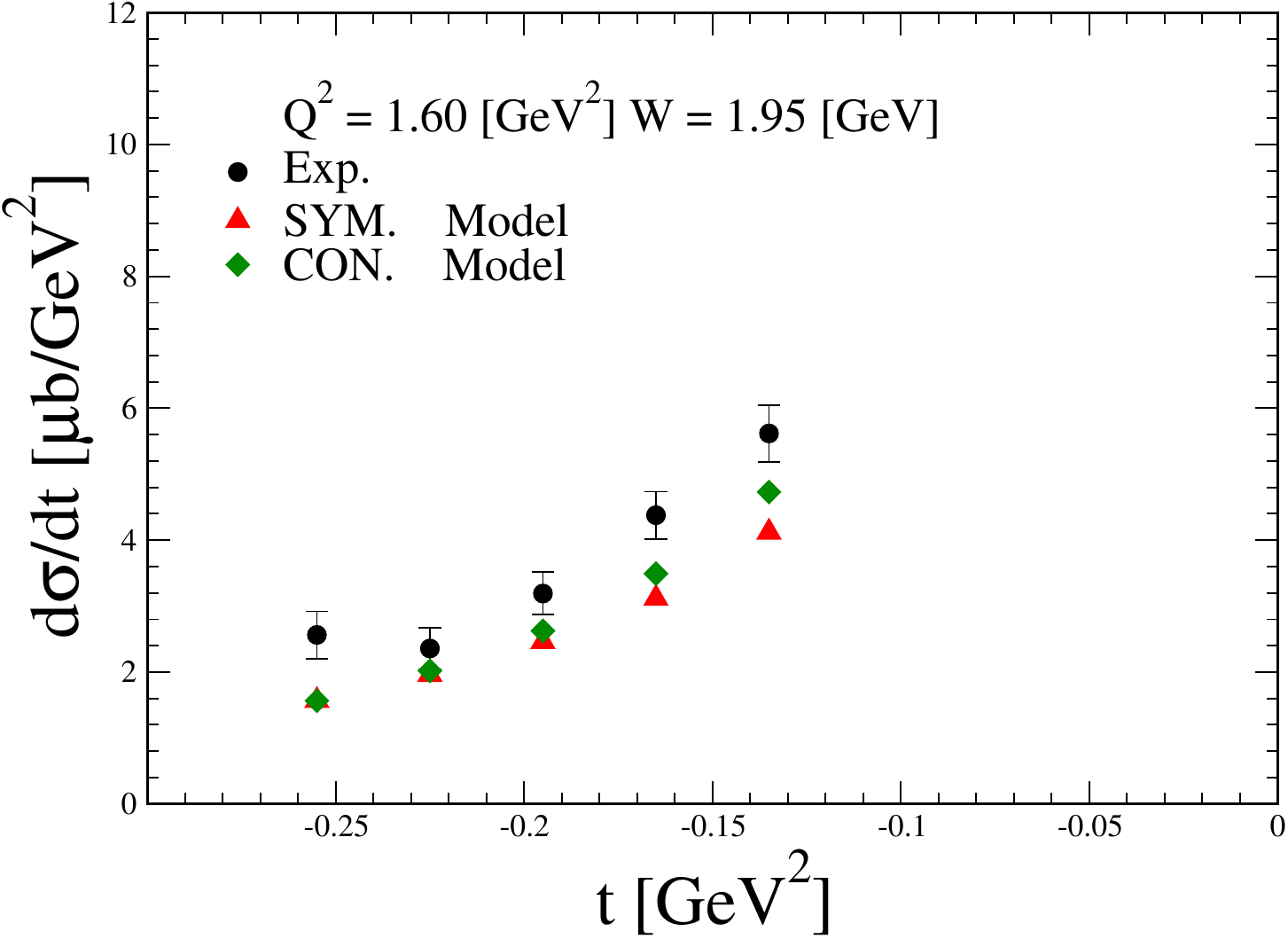,width=5.cm,angle=0}	
\epsfig{figure=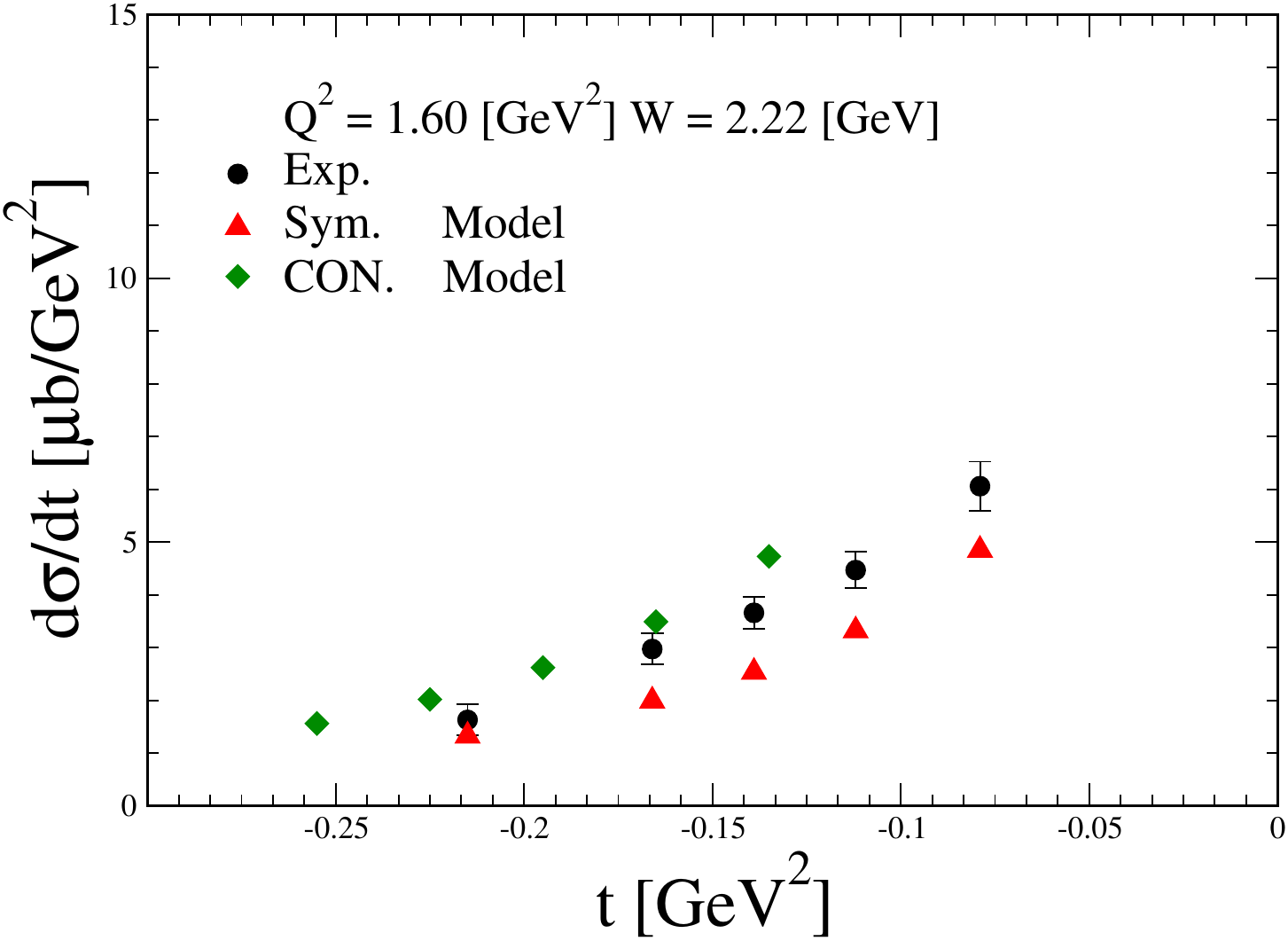,width=5.cm,angle=0}
\vspace{0.28cm}
\epsfig{figure=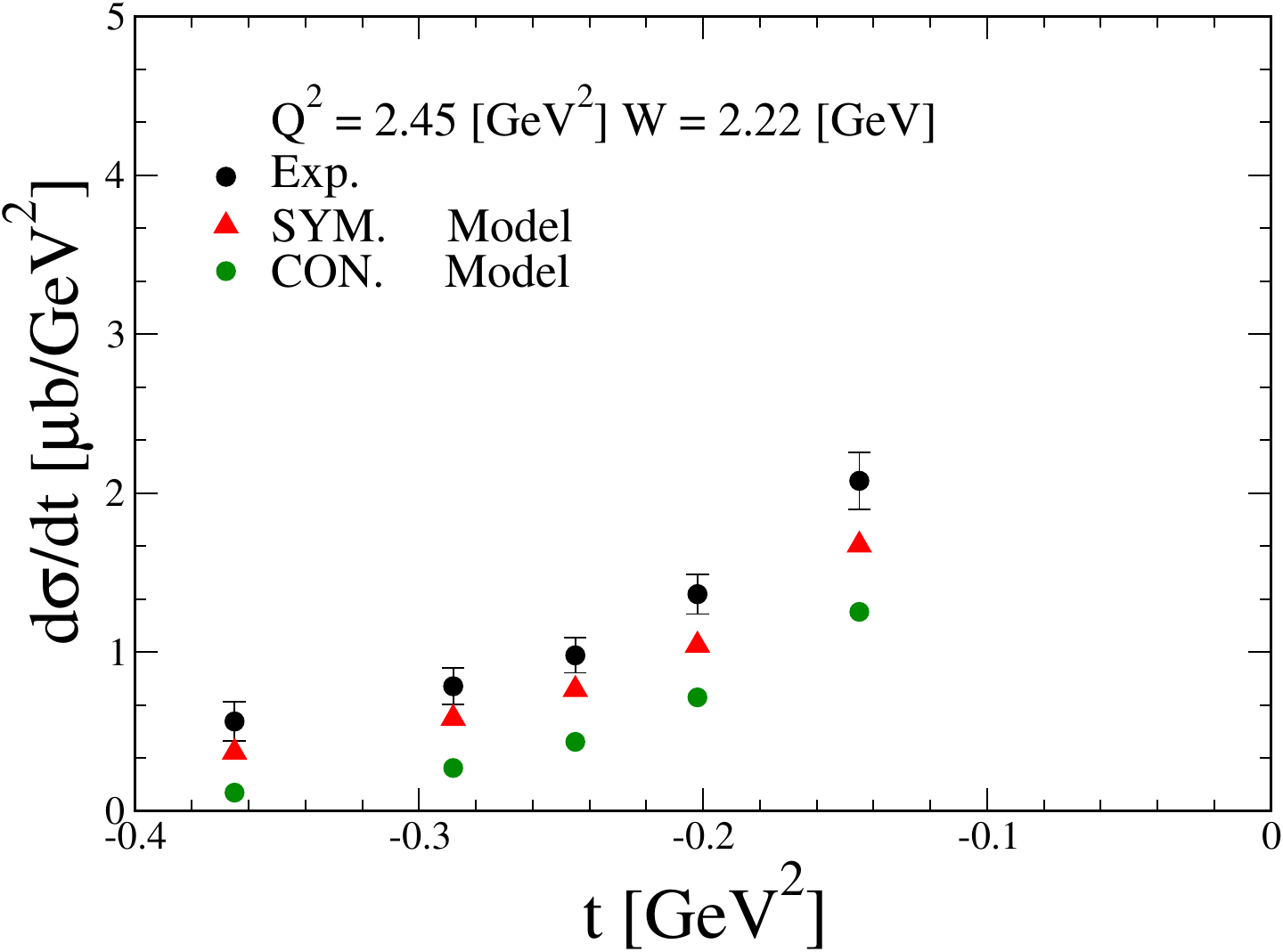,width=5.cm,angle=0} 
 \caption{Cross-sections calculated with CON and SYM models, compared with the experimental data, 
	for fixed $Q^2$ and $W$, and varying $t$. Experimental values and model results 
	are from Table~\ref{table3} presented in Appendix~\ref{app:tables}.
}
\label{Fig:crossections}
 \end{center} 
 \end{figure*}

\subsection{ Cross-section $d\sigma_{\rm L}/dt$}
\label{subsec:xsec}

 The experimental differential cross sections $d\sigma^{\rm Exp}_{\rm L}/dt$,
  for the exclusive pion photo-production on the nucleon from Ref.~\cite{Blok2008},
and  our model calculations $d\sigma^{\rm th}_{\rm L}/dt$ using the Chew-Low
 approach as described by Eq.~\eqref{eq:chew} 
 for the two different vertex models,
 $\Gamma_\pi=(\Gamma^{\rm CON}_\pi,\Gamma^{\rm SYM}_\pi)$,
are presented in Table~\ref{table3}.
This table is organized by the values of
$Q^2$, $W$ and $t$  as taken from the experimental data~\cite{Huber2008,Blok2008}.

The cross-section  $d\sigma_{\rm L}/dt$
as a function of $t$ is shown in Fig.~\ref{Fig:crossections} for
 both the experimental data and the results  from the two
  different models presented in Table~\ref{table3}. 
We plot $d\sigma_{\rm L}/dt$ for the six groups specified in Table~\ref{table3},
with average squared momentum transfers and 
energies $(\langle Q^2\rangle,\,W)$ in units of [GeV$^2$, GeV]
as follows:
(0.6,1.95),  (0.75,1.95), (1,1.95), (1.6,1.95), (1.6,2.22) and (2.45,2.22).
 As clearly seen in Fig.~\ref{Fig:crossections}, the model 
 dependence of the cross-section intensifies with higher $Q^2$ and $t$.
The  SYM (triangle)  model 
aligns closely with the experimental data for $Q^2$ up to 2.45 GeV$^2$. However,  the 
 constant vertex model (diamond) exhibits a larger deviation from the experimental cross-section data.

It is interesting to note that the model results 
for $d\sigma_{\rm L}/dt$ as a function of $t$ for fixed $Q^2$ 
 reflect the behavior of the off-shell form factor 
 $|F_1(Q^2,t)|^2$ with $t$. They are primarily influenced by the pion pole but only weakly dependent on $G_{\pi NN}(t)$ as expected.
This turns $d\sigma_{\rm L}/dt$  into an additional source of information 
about the pion charge distribution beyond its $Q^2$ dependence.  
Furthermore, our observations indicate that the SYM model demonstrates a consistency with the experimental data.
  This suggests that understanding their behavior with respect to $t$ can
   facilitate a more accurate determination of the off-shell form factor 
   and aid in extrapolating to the physical on-shell point.

While our previously studied CON model utilized dimensional regularization, the current SYM model employs
the Pauli-Villars regularization, which may achieve better convergence than dimensional regularization in the 
calculation of hadronic observables. As shown in Fig.~\ref{f1exp}, the results from the SYM model 
are indeed closer to the experimental data of the pion off-shell form factor $F_1 (Q^2,t)$ for the higher \( Q^2 \) and \( t \) ranges compared to the CON model. 
Further evidence of this improved agreement with the experimental data can be observed in Fig.~\ref{Fig:crossections}, 
where the differential cross sections in the SYM model show better alignment with the data for higher \( Q^2 \) and \( t \) ranges 
than those in the CON model.  
We attribute this enhanced description of the experimental data in these ranges to the better convergence properties of the quantum field theoretic model computation used in the SYM model.

\label{conclusions}
In this work, we  extended a previous study~\cite{Choi2019} 
by conducting a comparative analysis of the pion off-shell form factors, 
namely, $F_1(Q^2,t)$ and $g(Q^2,t)$. Our goal is to explore the model dependence of 
these observables within a constituent quark framework describing
 a  $q\bar q$ composite pion with massive yet structureless degrees of freedom.

In addition to the constant vertex model, which assumes
 a structureless pion-quark vertex~\cite{Choi2019}, 
this study utilizes the model that introduces a momentum 
scale to ensure the finiteness of the Mandelstam formula for
 the microscopic form factor. 
This model is employed to calculate the two off-shell form factors. 
Furthermore, we also extract the off-shell form factor 
 $g(Q^2,t)$ from the cross-section data~\cite{Blok2008},
 relying on the pion-pole dominance at small $t$ values
 in the Sullivan process  $^1$H$(e,e'\pi^+)n$. 
 The form  factor $g(Q^2,t)$ does not vanish when $t=m^2_\pi$,
 and its extraction from the data requires the knowledge of $F_1(0,t)$. 
 The calculation of $F_1(0, t)$ has been performed for both models
  utilized in this study.

These models of the pion, which incorporate constituent quark degrees of
freedom, reproduce the experimental charge radius with less than 10\% error. Notably,
the momentum scales of the regulators are specifically
chosen to reproduce the pion decay constant, which is 
	essential for accurately reproducing the pion charge radius.
The covariant model features a coupling $g_{\pi q\bar q}$ that is
	 specifically chosen
to better fit the pion charge radius and low energy on-shell form factor. 
Furthermore, it was established in the previous 
works~ \cite{deMelo1999,deMelo2002,Choi2019} that 
the resulting form factors $F_1(Q^2,m^2_\pi)$ are consistent with the data
below $Q^2=2$\,GeV$^2$ \cite{ Huber2008,Amen1,Amen2,Thorn1,Thorn2,Tadevosyan2007} 
within the reported errors, which we have presented  for reference in this work.
	 Building on this,
we computed $F_1(0,t)$  for $0<-t<0.4$~GeV$^2$ within the range of the experimental 
cross-section data for $d\sigma_{\rm L}/dt$ from the pion photo-production
process $^1$H$(e,e'\pi^+)n$ cited in Ref.~\cite{Blok2008}. This analysis
	 shows a moderate model dependence, which is
reflected in the extraction of $g(Q^2,t)$ over the data range 
 $0.526\leq Q^2 \leq 2.703\,$GeV$^2$ and $0.026\leq  -t\leq 0.365\,$GeV$^2$.
Furthermore, as $Q^2$ increases, the off-shell form factor $F_1(Q^2, t)$
 becomes more sensitive to the chosen model,  allowing for a more detailed
  resolution of the pion charge distribution.
The closely related form factor $g(Q^2,t)$ also exhibits 
mild model dependence below $Q^2=2$\,GeV$^2$, much of that associated with 
$g(0,m^2_\pi)=\la r^2_\pi\ra/6$. This value varies by less than 
10\% across the three models.

To demonstrate the usefulness of our work, we have presented
$F_1(Q^2,t)$, $g(Q^2,t)$ and $d\sigma_{\rm L}/dt$
 in Tables~\ref{table1}, \ref{table2} and \ref{table3}, respectively,
   each calculated using the two models as tabulated in Appendix~\ref{app:tables}. 
Additionally, we have included the  values 
 of $g(Q^2,t)$ extracted from the experimental cross-sections, using 
 	 the computed $F_1(0,t)$ as a basis. 
These calculations and the extracted values of $g(Q^2,t)$ can serve as a valuable
 reference for future investigations into the off-shell
 form factors of the pion. In particular, these studies at low values of $t$ 
and different $Q^2$ values can provide further insights
 into the charge distribution of the pion.
Moreover, these studies can be extended to include the kaon, exploring
 the momentum distributions of both the pion and kaon. Such investigations 
 are planned for future facilities, including the experimental 
 program at the Electron-Ion Collider (EIC)~\cite{AbdulKhalek2021,Burkert2022}.

It is worthwhile to stress that successful  extraction of the form factors
 from the cross-sections depend on the reliability
of the employed model.  It is now well established that dynamical
 chiral symmetry breaking occurs in the light flavor sector, 
where the pion and kaon act as Goldstone modes.  
This phenomenon is evident in both lattice QCD simulations (see e.g.~\cite{Falcao:2022gxt})
  and in continuum approaches~\cite{Roberts:2021nhw}.
  In these processes, the strong interaction modifies bare quarks by imparting a running mass, 
  which in models featuring structureless quarks, corresponds to the constituent mass. 
It may provide a reasoning for employing constituent quark degrees of freedom, as utilized in this work based on Refs.~\cite{deMelo1999,deMelo2002}, which 
has been successful in describing the characteristics of light mesons such as 
the pion. One may note that these models have successfully reproduced various pion properties,
 including the electromagnetic charge radius and the electroweak decay constant~\cite{pdg2020}.
  Additionally, when compared to experimental
   data~\cite{Blok2008,Huber2008,Dally1,Dally2,Amen1,Amen2,Thorn1,Thorn2}, 
   these models provide a reasonable approximation of the pion's electromagnetic 
  form factor to a certain extent.

Therefore, it is reasonable to anticipate that the use of simple models
would effectively be a step toward describing the pion off-shell form factors
 and cross-sections,
as indeed demonstrated in this work. 
 In addition to the models adopted  in this study,
  one may further explore incorporating dynamical models 
in Minkowski space~\cite{Ydrefors:2021dwa,dePaula:2022pcb} 
as the more advanced frameworks are necessary to integrate concepts 
such as running mass~\cite{Mezrag:2020iuo,Duarte:2022yur,Sauli:2022ild}
to provide a more comprehensive understanding.

\begin{acknowledgements}

This work was partially supported by Funda\c c\~ao de Amparo \`a Pesquisa 
do Estado de S\~ao Paulo (FAPESP)  grant 2019/07767-1 (TF), 
2019/02923-5 and  2023/09539-1 (JPBCM), and by Conselho Nacional de Desenvolvimento Cient\'\i fico 
e Tecnol\'ogico (CNPq) grants 306834/2022-7(TF), 307131/2020-3 (JPBCM), 
and 464898/2014-5 (INCT-FNA).
The work of H.-M.C. was supported by the National 
Research Foundation of Korea (NRF) under Grant No. NRF- 2023R1A2C1004098.
The work of C.-R.J. was supported in part by the U.S. Department of Energy 
(Grant No. DE-FG02-03ER41260). 
The National Energy Research Scientific Computing Center (NERSC)
 supported by the Office of Science of the U.S. Department of Energy 
under Contract No. DE-AC02-05CH11231 is also acknowledged. 
\end{acknowledgements}

\appendix{}

\section{ $k^-$ integration}
\label{app:kmint}

\begin{figure}[b]
\begin{center}
\includegraphics[height=4cm,width=8cm] {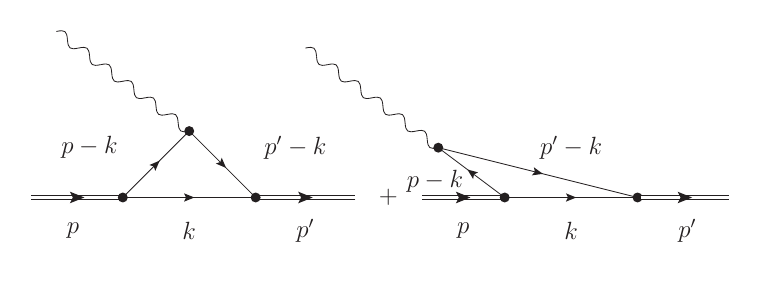}
\vspace{-0.5cm}
\caption{Diagrammatic representation of the valence and non-valence contributions
			to the pion  photo-absorption amplitude obtained after $k^-$ integration. 
			Left diagram corresponds to the valence  region $0<k^+<p^+$ in the loop integral,
			and the right one corresponds to  the non-valence 
			region  $p^+<k^+<p^++q^+=p^{\prime +}$ in the loop integral.} 
\label{triangle2}
\end{center}
\end{figure}

In this appendix, we outline the use of the light-front variables 
to perform the loop-integration
of Eq.~\eqref{eq:Mandpioncurr} for the plus component of the pion current,
which facilitates the extraction of the form factors.
We allow $q^+>0$ in 
order to separate the valence and non-valence contributions, 
as illustrated in Fig.~\ref{triangle2} by the left and right
diagrams, respectively. We will demonstrate that 
in the limit $q^+\to 0_+$, only the valence contribution remains, 
while the non-valence contribution to the 
loop integral vanishes (see e.g. Refs.~\cite{deMelo1998,deMelo1999}).
Consequently, in the model adopted, there are
no zero-mode contributions to the pion EM form factor,
 regardless of whether $p^2<0$ or $p^2=m^ 2_\pi$ and $p^{\prime 2}=m^2_{\pi}$.

The matrix element of Eq.~\eqref{eq:Mandpioncurr} 
can be computed from using the plus component of the current as:
\begin{multline}
\Gamma^{+({\rm SYM})} = 
\int \frac{d^2k_\perp dk^+ dk^-}{2 (2 \pi)^4} 
\frac{i \,C_1\,\text{Tr}[ {\cal O}^+]}{[1][2][3]}
\\ \times
\left[ \frac{1}{[4]} + \frac{1}{[5]} \right] 
\left[ \frac{1}{[6]} + \frac{1}{[7]} \right]~.
\label{eq:inte2}
\end{multline}
Here, $C_1=-2 e\dfrac{m^2}{f_\pi^2} N_c N^{2}$ and 
\begin{equation}
{\cal O}^+= (\rlap\slash k +m)\gamma^5(\rlap\slash k - \rlap\slash p^\prime +m)
\gamma^+ (\rlap\slash k - \rlap\slash p + m) \gamma^5 \, ,
\label{O}
\end{equation} 
with the trace computed as:
\begin{eqnarray}
\text{Tr}[{\cal O}^+ ] =  &-&   4\,k^-(p^{\prime +}-k^+)(p^{+}-{k^+})
\nonumber\\
&+&4\,(k^2_\perp+m^2)(k^+-p^+-p^{\prime +})
\nonumber \\
&-& 2\, \vec k_\perp \cdot 
\vec q_\perp
\, q^++ k^+q^2_\perp
\,.
\label{eq:tr}
\end{eqnarray}
The denominators in Eqs.~\eqref{eq:inte2}  
are given by
\begin{eqnarray} \nonumber
\left[ 1 \right] & = & k^+ \left[ k^- -
\frac{f_1-\imath \epsilon}{k^+}  \right],
\\ \nonumber 
\left[ 2 \right] & = &  
\left( p^+ -k^+ \right)  \left[p^- - k^- - 
\frac{f_2-\imath \epsilon}{p^+-k^+} \right],
\\ \nonumber
\left[ 3 \right] & = & 
\left(p^{\prime +} - k^+ \right) \left[p^{\prime -} -
k^- - \frac{f_3-\imath \epsilon}{p^{\prime +}-k^+}  \right],
\\ \nonumber
\left[ 4 \right] & = & k^+ \left[ k^- - \frac{f_4-\imath \epsilon}{k^+}  \right],
\\ \nonumber
\left[ 5 \right] & = & 
\left( p^+ - k^+ \right) \left[p^- - k^- - \frac{f_5-\imath \epsilon}{p^+-k^+}  \right],
\\  \nonumber
\left[ 6 \right] & = & k^+ \left[ k^- -
\frac{f_6-\imath \epsilon}{k^+}  \right]\, ,  \\ 
\left[ 7\right] & = & 
\left( p^{\prime +} -k^+ \right)
\left[p^{\prime -} - k^- - \frac{f_7-\imath \epsilon}{p^{\prime+} - k^+}  \right],
\end{eqnarray}
In the expressions for the denominators above, we have the following 
definitions of $f_j(j=1,\cdots,7)$: 
\begin{eqnarray}
&&f_1=k^2_\perp+m^2\,, \quad f_2=(p-k)^2_\perp+m^2\, ,
\nonumber \\
&&f_3=(p^\prime-k)^2_\perp+m^2\,,\quad 
f_4=k^2_\perp+m_R^2\,, 
\nonumber \\
&& f_5=(p-k)^2_\perp+m^2_R\,,\quad f_6=k^2_\perp+m_R^2\, ,
\nonumber \\ && f_7=(p^{\prime} - k)^2_\perp+m^2_R\,.
\end{eqnarray}

The Cauchy integral on $k^-$ for Eqs.~\eqref{eq:inte2}, 
is novanishing only in the intervals:
(i) $0 < k^+ <p^+$ in the valence region, and 
(ii) $p^+ < k^+ <p^{\prime +}$ in the non-valence region, when $q^ +>0$.
The remaining question concerns what happens when $q^+\to0_+$, 
where one might naively expect no zero-mode contributions. We will 
demonstrate that this is indeed the case in the adopted pion models,
where only the valence region contributes to the photo-absorption amplitude.

Let us begin by discussing Cauchy integration in the valence region where $0<k^+<p^+$.
By  adopting the integration path as described 
in Eq.~\eqref{eq:inte2} for the 
SYM model,  the residues must be evaluated at the poles
$k^-_i$ for $i=\,$1, 4 and 6, resulting
in a lengthy but straightforward expression. 
The resulting formulas for the valence contributions to 
$\Gamma^{+({\rm SYM})}(p',p)$, with $p^{\prime 2}=m^2_\pi$
and $p^2<0$ or equal to $m^2_\pi$, 
exhibit no further singularities under the  condition
$m_\pi<2m<m+m_R<2m_R$ set by our model parametrization. 
Therefore,  they are amenable to numerical integration over $k^+$ and $\vec k_\perp$.

The non-valence (referred to as ``$nval$") contribution to 
\eqref{eq:inte2} is obtained by closing the integration
path in the upper complex plane and computing 
the residues at the poles: 
\begin{equation}\label{eq:kjnval}
k^-_{j\,nval}= p^{\prime - } - \frac{f_j - \imath \epsilon }{p^{\prime +}-k^+}
\end{equation}
with $j=\,$3 and 7. However, for the sake of simplicity, we only present
the results from the pole $k^-_3$, particularly in the limit as $q^+\to0+$.
This pole is located on the upper complex semi-plane when  $p^+<k^+<p^{\prime +}$. 
To analyze the limit approaching the Drell-Yan frame,  where $q^+\to0_+$,
we express $k^+$ as
$k^+=x q^+ + p^+$ in terms of the new variable $0<x<1$.  Therefore, 
\begin{equation}\label{eq:k3nval}
k^-_{3\, nval}\sim-\frac{f_3}{q^+x} \Big|_{q^+\to 0+}.
\end{equation}
We then count the lowest power in $q^+$, noting that the trace in 
Eq.~\eqref{eq:tr} is of $\mathcal{O}\left[(q^+)^0]\right)$.

For the SYM model, the Cauchy integral of Eq.~\eqref{eq:inte2} involves
considering the residue at
the pole $k^-_{3\,nval}$  as given in Eq.~\eqref{eq:k3nval}.  In addition,
we must also 
consider the contributions from:
\begin{equation}\label{eq:orderden2}
[j]^{-1}\sim \mathcal{O}[q^+]\,,\quad
j=4~\text{and}~6 \, .
\end{equation}
These two additional denominators produces subleading contributions 
of $\mathcal{O}[(q^{+})^2$
and $\mathcal{O}[(q^{+})^3]$ leaving only the dominant 
leading order  term for $q^+\to 0_+$:
\begin{equation}\label{eq:symnval}
\Gamma^{+({\rm SYM})}_{nval} \sim \mathcal{O}[q^+]  ~.
\end{equation} 
The Formulas analogous to Eqs.~\eqref{eq:symnval},    
are derived for the residue contribution from the pole $k^-_{7\,nval}$ 
as specified in Eq.~\eqref{eq:kjnval}. 
It is worth mentioning that similar reasoning 
was previously used to find the contributions of the LF zero-modes
to the EM current of a composite fermion-antifermion 
spin-1 particle~\cite{deMelo:2012hj}.

The limit $q^+\to 0_+$ is  readily applied to Eq.~\eqref{eq:symnval},      
 which are proportional to $q^+$:   
\begin{equation}
\lim_{q^+ \rightarrow 0_+}  \Gamma^{+({\rm SYM})}_{nval} ~ =  0~,
\end{equation}
indicating that, even in the off-shell regime with $p^2<0$,  the plus
component of the pion electromagnetic current in this model does not acquire
non-valence  or zero mode contributions in the Drell-Yan frame. Therefore,  our numerical 
calculations  can focus exclusively on the valence region to obtain the form factors.

For an extension to unequal mass system such as the kaon, one may replace 
Eq.\eqref{eq:inte2} by  
\begin{equation}
		\Gamma^{+({\rm NOSY})} =  
	\int \frac{d^2k_\perp dk^+ dk^-}{2 (2 \pi)^4} 
	\frac{i\, C_1\,\text{Tr}[ {\cal O}^+]}{[1][2][3][5][7]}.
	\label{eq:inte1}
\end{equation}
For the NOSY model in Eq.~\eqref{eq:inte1}, 
the residue is derived solely from the pole:  
\begin{equation}
k^-_{j\,val}=\frac{f_j-\imath \epsilon}{k^+}  
\end{equation}
with $j=1$, when the integration path is closed in the lower
complex semi-plane. Analyzing  Eq.~\eqref{eq:inte1} and  focusing on the residue at the 
pole  $k^-_{3\, nval}$, the residue is computed from the product of 
the denominators $[j]$ with $j=\,$1, 2,  5 and 7.  By counting
the powers of $q^+$ in these denominators, we find that:
\begin{equation}
\label{eq:orderden1}
[j]^{-1}\sim \mathcal{O}[(q^+)^{n }]\,,\quad
n=\delta_{j,1} \, ,
\end{equation}
which indicates that the contribution of each denominator scales 
according to the Kronecker delta function $\delta_{j,1}$, 
primarily affecting the first denominator. Collecting these results,
we can find that:
\begin{equation}  
\Gamma^{+({\rm NOSY})}_{nval}\sim  \mathcal{O}[q^+]
\, . \label{eq:nosynval}
\end{equation} 
This shows that the non-valence contribution of $\Gamma^{+({\rm NOSY})}$ 
scales with $q^+$ as in the case of $\Gamma^{+({\rm SYM})}$.
Even in the off-shell regime with $p^2<0$, thus, the plus
component of the pion electromagnetic current in both 
 SYM and NOSY models does not acquire non-valence  or zero mode contributions in the Drell-Yan frame.

\begin{widetext}

\section{Tables of off-shell form factors and cross-sections}\label{app:tables}

This appendix is devoted to presenting the computed values of the pion form factors, $F_1(Q^2,t)$ and $g(Q^2,t)$,
shown in Tables~\ref{table1} and~\ref{table2}, respectively. Theses values are derived from the two models given by  Eqs.~\eqref{eq:sym} and \eqref{eq:cov},  which are used to construct the microscopic pion EM current. 
The experimentally extracted $g(Q^2,t)$, shown in Table~\ref{table2}, requires the theoretical values of $F_1(0,t)$ from Table~\ref{table1}
and the extracted  $F_1(Q^2,t)$~\cite{Choi2019} from the exclusive experimental cross-sections~\cite{Blok2008}.
These values of $d\sigma_L/dt$ for the exclusive pion photo-production process on the nucleon, along with calculated values
from Eq.~\eqref{eq:chew} for the three models, are presented in Table~\ref{table3}.

\begin{table}[h]
	\centering
	\caption{Electromagnetic form factors $F_1(Q^2,t)$ and $F_1(0,t)$ 
			 from the CON and SYM models.  
	The 	experimentally extracted form factor $F^{\rm Exp}_1(Q^2,t)$
	 is tabulated in the fourth column.
	}
	\label{table1}
	\begin{tabular}{ccc|ccc|cc}
		\hline 
		\hline
$Q^2$ & $W$  & $-t$  & \multicolumn{3}{c|}{$F_1(Q^2,t)$} & \multicolumn{2}{c}{$F_1(0,t)$} \\
		\multicolumn{1}{l}{$({\rm GeV}^2)$} & \multicolumn{1}{l}{$({\rm GeV})$} & \multicolumn{1}{l|}{$({\rm GeV}^2)$} 
                    & Exp.  & CON     & SYM          & CON    & SYM      \\ 
\hline   \hline 
	  &   &  &  &     &    &        &   
	 \\
  0.526  & 1.983  & 0.026 & 0.502$\pm$0.013 & 0.487    & 0.4471        & 0.891   & 0.8926  
	 \\
	0.576  & 1.956  & 0.038 & 0.440$\pm$0.010 & 0.462    & 0.4200       & 0.869   & 0.8685       \\
	0.612  & 1.942  & 0.050 & 0.413$\pm$0.011 & 0.443    & 0.3995       & 0.849   & 0.8458     \\
	0.631  & 1.934  & 0.062 & 0.371$\pm$0.014 &  0.430    & 0.3860      & 0.831   & 0.8244      \\
	0.646  & 1.929  & 0.074 & 0.340$\pm$0.022 &  0.419    & 0.3744      & 0.814   & 0.8041       \\
	&      &        &       &                &            &              &                     \\
	0.660  & 1.992  & 0.037  &  0.397$\pm0.019$  &  0.435    & 0.3937     & 0.870   & 0.8705      \\
	0.707  & 1.961  & 0.051  & 0.360$\pm$0.017   & 0.414    & 0.3717         & 0.848   & 0.8440      \\
	0.753  & 1.943  & 0.065  & 0.358$\pm$0.015   & 0.394    & 0.3526        & 0.827   & 0.8192       \\
	0.781  & 1.930  & 0.079  & 0.324$\pm$0.018   & 0.381    & 0.3382       & 0.807   & 0.7960       \\
	0.794  & 1.926  & 0.093  & 0.325$\pm$0.022   & 0.371    & 0.3289         & 0.789   & 0.7742      \\
    &      &        &        &          &           &            &                          \\
	0.877  & 1.999  & 0.060  &  0.342$\pm$0.014   & 0.366    & 0.3283        & 0.834   & 0.8279      \\
	0.945  & 1.970  & 0.080  &  0.327$\pm$0.012   & 0.343    & 0.3058        & 0.806   & 0.7944      \\
	1.010  & 1.943  & 0.100  & 0.311$\pm$0.012    & 0.322    & 0.2868        & 0.781   & 0.7638      \\
	1.050  & 1.926  & 0.120  & 0.282$\pm$0.016    & 0.307    & 0.2731        & 0.758   & 0.7357      \\
	1.067  & 1.921  & 0.140  & 0.233$\pm$0.028    & 0.297    & 0.2637        & 0.737   & 0.7097      \\
		   &        &        &             &           &            &                        \\
	1.455  & 2.001  & 0.135  & 0.258$\pm$0.010  & 0.237    & 0.2227        & 0.742   & 0.7160      \\
	1.532  & 1.975  & 0.165  & 0.245$\pm$0.010  & 0.219    & 0.2078        & 0.714   & 0.6799      \\
	1.610  & 1.944  & 0.195  &  0.222$\pm$0.012 & 0.201    & 0.1955        & 0.688   & 0.6475      \\
	1.664  & 1.924  & 0.225  & 0.203$\pm$0.013  & 0.188    & 0.1860        & 0.665   & 0.6182      \\
	1.702  & 1.911  & 0.255  & 0.227$\pm$0.016  & 0.177    & 0.1783       & 0.644   & 0.5896      \\
	       &        &        &                   &           &                 &                 \\
	1.416  & 2.274  & 0.079  &  0.270$\pm$0.010  & 0.259    & 0.2430        & 0.807   & 0.7945       \\
	1.513  & 2.242  & 0.112  & 0.258$\pm$0.010   & 0.235    & 0.2238        & 0.767   & 0.7450       \\
	1.593  & 2.213  & 0.139  &0.251$\pm$0.010    &  0.217    & 0.2097       & 0.738   & 0.7092       \\
	1.667  & 2.187  & 0.166  & 0.241$\pm$0.012   & 0.201    & 0.1976        & 0.713   & 0.6769       \\
	1.763  & 2.153  & 0.215  & 0.200$\pm$0.018  &  0.179    & 0.1816        & 0.672   & 0.6257       \\
		   &        &        &            &           &            &                       \\
	2.215  & 2.308  & 0.145  & 0.188$\pm$0.008  & 0.146    & 0.1691        & 0.732   & 0.7017     \\
	2.279  & 2.264  & 0.202  &  0.178$\pm$0.008 & 0.129    & 0.1570        & 0.682   & 0.6385      \\
	2.411  & 2.223  & 0.245  & 0.163$\pm$0.009 & 0.109    & 0.1457         & 0.650   & 0.5982     \\
	2.539  & 2.181  & 0.288  & 0.156$\pm$0.011 & 0.092    & 0.1352         & 0.622   & 0.5630     \\
    2.703  & 2.127  & 0.365  & 0.150$\pm$0.016 & 0.068    & 0.1223         & 0.579   & 0.5097     \\ 	
      &   &  &  &     &    &      &       \\
\hline 
\hline
\end{tabular}
\end{table}

\begin{table}[h]
	\centering
	\caption{
		The form factor $g(Q^2,t)$ (in units of [GeV]$^{-2}$),
		 with $g^{\rm CON}$ and $g^{\rm SYM}$ extracted
		 from the experimental cross-sections 
		 in Table VII of Ref.~\cite{Blok2008}, 
		 through $F_1^{\rm Exp}(Q^2,t)$ and $F_1(0,t)$ 
		 (see Table~\ref{table1}) from the CON and SYM models, respectively.
		The form factors $g_{th}^{{\rm CON}}$ and $g_{th}^{{\rm SYM}}$ 
		are the theoretical results calculated with the two models.	
	}
	\label{table2}
	\begin{tabular}{cccccc}
		\hline 
		\hline
		$Q^2~[{\rm GeV}^2]$ & $-t~[{\rm GeV}^2]$ & $g_{th}^{{\rm CON}}$  & $g_{th}^{{\rm SYM}}$ 
		 & $g^{\text{CON}}(Q^2,t)$ 		& $g^{\text{SYM}}(Q^2,t)$
		\\ 
		\hline 
		\hline
		&       &                      &                      &                                  \\
		0.526 & 0.026 & 0.768 & 0.846     & $0.7401 \pm 0.046$      & $0.7429 \pm 0.046$     \\
		0.576 & 0.038 & 0.708 & 0.779     & $0.7442 \pm 0.031$      & $0.7434 \pm 0.030$      \\
		0.612 & 0.050 & 0.664 & 0.727    & $0.7122 \pm 0.085$      & $0.7072 \pm 0.029$    \\
		0.631 & 0.062 & 0.635 & 0.693    & $0.7284 \pm 0.036$      & $0.7177 \pm 0.036$     \\
		0.646 & 0.074 & 0.611 & 0.665    & $0.7343 \pm 0.053$      & $0.7188 \pm 0.053$      \\  
		&       &                        &                      &                              \\
		0.660 & 0.037 & 0.660 & 0.722    & $0.7166 \pm 0.044$      & $0.7170 \pm 0.044$     \\
		0.707 & 0.051 &  0.613 & 0.668   & $0.6900 \pm 0.033$      & $0.6843 \pm 0.033$     \\
		0.753 & 0.065 & 0.574 & 0.620    & $0.6224 \pm 0.027$      & $0.6105 \pm 0.027$   \\
		0.781 & 0.079 & 0.546 & 0.584    & $0.6182 \pm 0.030$      & $0.6041 \pm 0.030$     \\
		0.794 & 0.093 & 0.526 & 0.560    & $0.5845\pm0.035$      & $0.5631 \pm 0.036$       \\
		&       &                        &                         &                               \\
		0.877 & 0.060 & 0.533 & 0.570    & $0.5615 \pm 0.018$      & $0.5535 \pm 0.018$    \\
		0.945 & 0.080 & 0.490 & 0.512    & $0.5067 \pm 0.014$      & $0.4940 \pm 0.014$   \\
		1.010 & 0.100 & 0.454  & 0.472   & $0.4657 \pm 0.012$      & $0.4488 \pm 0.012$  \\
		1.050 & 0.120 & 0.430 & 0.440    & $0.4537 \pm 0.014$      & $0.4308 \pm 0.014$   \\
		1.067 & 0.140 & 0.412 & 0.418    & $0.4726 \pm 0.025$  & $0.4464 \pm 0.025$       \\
		&       &                        &                         &                   
		\\
		1.455 & 0.135 & 0.347 & 0.339    & $0.3324 \pm 0.005$   & $0.3145 \pm 0.005$     \\
		1.532 & 0.165 & 0.323 & 0.308    & $0.3059 \pm 0.004$      & $0.2837 \pm 0.004$   \\
		1.610 & 0.195 & 0.302 & 0.280    & $0.2896 \pm 0.004$      & $0.2645 \pm 0.004$     \\
		1.664 & 0.225 & 0.286 & 0.259    & $0.2775 \pm 0.005$      & $0.2494 \pm 0.005$     \\
		1.702 & 0.255 & 0.274 & 0.242    & $0.2452 \pm 0.005$      & $0.2144 \pm 0.005$      \\
		&       &                        &                         &                             \\
		1.416 & 0.079 & 0.387 & 0.300    & $0.3792 \pm 0.005$      & $0.3714 \pm 0.005$      \\
		1.513 & 0.112 & 0.351 & 0.344    & $0.3362 \pm 0.004$      & $0.3226 \pm 0.004$      \\
		1.593 & 0.139 & 0.327 & 0.313    & $0.3058 \pm 0.004$      & $0.2889 \pm 0.004$      \\
		1.667 & 0.166 & 0.307 & 0.288    & $0.2831 \pm 0.004$      & $0.2626 \pm 0.004$    \\
		1.763 & 0.215 & 0.280 & 0.251    & $0.2676 \pm 0.005$      & $0.2425 \pm 0.006$     \\
		&       &                        &                         &                           \\
		2.215 & 0.145 & 0.265 & 0.240    & $0.2456 \pm 0.002$      & $0.2327 \pm 0.002$     \\
		2.279 & 0.202 & 0.243 & 0.212    & $0.2211 \pm 0.002$      & $0.2028 \pm 0.002$     \\
		2.411 & 0.245 & 0.224 & 0.187    & $0.2018 \pm 0.002$      & $0.1811 \pm 0.002$      \\
		2.539 & 0.288 & 0.209 & 0.167    & $0.2834 \pm 0.002$      & $0.1607 \pm 0.002$       \\
		2.703 & 0.365 & 0.189 & 0.144    & $0.1589 \pm 0.002$      & $0.1337 \pm 0.002$    \\
		&     &                          &                         &                            \\
		\hline
		\hline
	\end{tabular}
\end{table}

\begin{table}[h]
	\centering
	\caption{
		Differential cross section~$d\sigma_L/dt$ 
		from the CON and SYM models compared to the experimental data~\cite{Blok2008}.
	}
	\label{table3}
	\begin{tabular}{ccc|ccc}
		\hline\hline
		$Q^2$           & $W$                & $-t$                  & \multicolumn{3}{c}{$d\sigma_L/dt~[\mu b/\text{GeV}^2]$}   \\
		$[\text{GeV}^2]$ & $[\text{GeV}]$ & $[\text{GeV}^2]$ & Exp    & CON    & SYM    
		\\
		\hline \hline
  &                    &                      &                    &        &                   \\
0.526                & 1.983              & 0.026                & 31.360 &  29.547   & 24.902   \\
0.576                & 1.956              & 0.038                & 24.410 &  26.874   & 22.208   \\
0.612                & 1.942              & 0.050                & 20.240 &  23.276   & 18.926   \\
0.631                & 1.934              & 0.062                & 14.870 &  19.932   & 16.061   \\
0.646                & 1.929              & 0.074                & 11.230 &  17.091   & 13.645   \\
&                    &                      &                    &        &                      \\
0.660                & 1.992              & 0.037                & 20.600 &  24.728   & 20.184   \\
0.707                & 1.961              & 0.051                & 16.280 &  21.509   & 17.338   \\
0.753                & 1.943              & 0.065                & 14.990 &  18.130   & 14.518   \\
0.781                & 1.930              & 0.079                & 11.170 &  15.4250  & 12.155   \\
0.794                & 1.926              & 0.093                & 9.949  &  12.9734  & 10.198   \\
&                    &                      &                    &        &             \\
0.877                & 1.999              & 0.060                & 14.280 &  16.398   & 13.195   \\
0.945                & 1.970              & 0.080                & 11.840 &  13.017   & 10.348   \\
1.010                & 1.943              & 0.100                & 9.732  &  10.453   & 8.295     \\
1.050                & 1.926              & 0.120                & 7.116  &  8.455    & 6.691    \\
1.067                & 1.921              & 0.140                & 4.207  &  6.852    & 5.404    \\
&                    &                      &                    &        &                       \\
1.455                & 2.001              & 0.135                & 5.618  &  4.729    & 4.175     \\
1.532                & 1.975              & 0.165                & 4.378  &  3.492    & 3.144    \\
1.610                & 1.944              & 0.195                & 3.191  &  2.624    & 2.483    \\
1.664                & 1.924              & 0.225                & 2.357  &  2.019    & 1.976    \\
1.702                & 1.911              & 0.255                & 2.563  &  1.563    & 1.587    \\
&                    &                    &                      &        &                       \\
1.416                & 2.274              & 0.079                & 6.060  &  5.575    & 4.908     \\
1.513                & 2.242              & 0.112                & 4.470  &  3.702    & 3.357     \\
1.593                & 2.213              & 0.139                & 3.661  &  2.740    & 2.561     \\
1.667                & 2.187              & 0.166                & 2.975  &  2.068    & 2.001     \\
1.763                & 2.153              & 0.215                & 1.630  &  1.303    & 1.341    \\
&                    &                      &                    &        &                \\
2.215                & 2.308              & 0.145                & 2.078  &  1.2521   & 1.681     \\
2.279                & 2.264              & 0.202                & 1.365  &  0.714    & 1.059     \\
2.411                & 2.223              & 0.245                & 0.980  &  0.436    & 0.779     \\
2.539                & 2.181              & 0.288                & 0.786  &  0.272    & 0.587     \\
2.703                & 2.127              & 0.365                & 0.564  &  0.116    & 0.377     \\
&                    &                      &                    &        &                        \\
\hline
\hline
\end{tabular}
\end{table}

\end{widetext}

\clearpage 
\newpage 


\end{document}